\documentclass{aa}
\usepackage{txfonts}
\usepackage{float}
\usepackage{graphicx}
\usepackage{natbib}
\usepackage{multirow}
\usepackage{wasysym} 
\bibpunct{(}{)}{;}{a}{}{,}

\usepackage{color}

\newcommand{\kms}{\,{\rm km\,s}^{-1}}

\newcommand{\Msun}{\,\mathrm{M}_\odot}

\newcommand{\simless}{\mathbin{\lower 3pt\hbox
      {$\rlap{\raise 5pt\hbox{$\char'074$}}\mathchar"7218$}}} 
\newcommand{\simgreat}{\mathbin{\lower 3pt\hbox
     {$\rlap{\raise 5pt\hbox{$\char'076$}}\mathchar"7218$}}} 

\newcommand{\msun}{{\,\rm M}_{\odot}}

\newcommand{\nht}{\ifmmode {{\rm NH}_3} \else {NH{\bas 3}} \fi}
\newcommand{\tco}{\ifmmode {^{13}{\rm CO}} \else {$^{13}{\rm CO}$}\fi}
\newcommand{\dco}{\ifmmode {^{12}{\rm CO}} \else {$^{12}{\rm CO}$}\fi}
\newcommand{\cdo}{\ifmmode {{\rm C}^{18}{\rm O}} \else {${\rm C}^{18}{\rm O}$}\fi}

\newcommand{\htco}{\ifmmode {{\rm H}^{13}{\rm CO}^{+} } \else {${\rm H}^{13}
{\rm CO}^{+}$ }\fi}
\newcommand{\hco}{\ifmmode {{\rm H}^{12}{\rm CO}^{+} } \else {${\rm H}^{12}
{\rm CO}^{+}$ }\fi}
\newcommand{\juz}{\ifmmode {{\rm J}=1\rightarrow 0} \else
{J=1$\rightarrow$0}\fi}
\newcommand{\jdu}{\ifmmode {{\rm J}=2\rightarrow 1} \else
{J=2$\rightarrow$1}\fi}
\newcommand{\jtd}{\ifmmode {{\rm J}=3\!\rightarrow\!2} \else
{${\rm J}=3\!\rightarrow\!2$} \fi}
\newcommand{\jcq}{\ifmmode {{\rm J}=5\!\rightarrow\!4} \else
{${\rm J}=5\!\rightarrow\!4$} \fi}
\newcommand{\jsc}{\ifmmode {{\rm J}=6\!\rightarrow\!5} \else
{${\rm J}=6\!\rightarrow\!5$} \fi}
\newcommand{\jss}{\ifmmode {{\rm J}=7\!\rightarrow\!6} \else
{${\rm J}=7\!\rightarrow\!6$} \fi}
\newcommand{\as}{\ifmmode {^{\scriptscriptstyle\prime\prime}}
        \else $^{\scriptscriptstyle\prime\prime}$\fi}
\newcommand{\am}{\ifmmode {^{\scriptscriptstyle\prime}}
        \else $^{\scriptscriptstyle\prime}$\fi}
\begin{document}
\title{GG Tau A: gas properties and dynamics from the cavity to the outer disk}
 \author {
	N.T. Phuong \inst{1^\star,2,3}	
        \and A. Dutrey \inst{1}
        \and P. N. Diep \inst{2,3}
        \and S. Guilloteau \inst{1}
   	\and E. Chapillon \inst{1,4}
	\and E. Di Folco  \inst{1} 
	\and Y-W. Tang \inst{5}
	\and V. Pi\'etu \inst{4}
	\and J. Bary \inst{6}
	\and T. Beck \inst{7}
	\and F. Hersant \inst{1}
	\and D.T Hoai \inst{2}
	\and J.M. Hur\'e \inst{1}
	\and P.T. Nhung \inst{2}
	\and A. Pierens \inst{1}
	\and P. Tuan-Anh \inst{2}
	 }      
 \institute{
		Laboratoire d'Astrophysique de Bordeaux, Universit\'e de Bordeaux, CNRS, B18N, 
			  All\'ee Geoffroy Saint-Hilaire, F-33615 Pessac; $^\star$ thi-phuong.nguyen@u-bordeaux.fr 
		 \and Department of Astrophysics, Vietnam National Space Center, 
			  Vietnam Academy of Science and Techonology, 18 Hoang Quoc Viet, Cau Giay, Hanoi, Vietnam 
  		 \and Graduate University of Science and Technology, Vietnam Academy of Science and Techonology, 18 Hoang Quoc Viet, Cau Giay, Hanoi, Vietnam
		 \and IRAM, 300 rue de la piscine, F-38406 Saint Martin d'H\`eres Cedex, France
		  \and Academia Sinica Institute of Astronomy and Astrophysics, PO Box 23-141, Taipei 106, Taiwan
		 \and Department of Physics and Astronomy, Colgate University, 13 Oak Drive, Hamilton, New York 13346, USA
		 \and Space Telescope Science Institute, 3700 San Martin Drive, Baltimore, Maryland 21218, USA}

\offprints{Nguyen Thi Phuong, \\
\email{thi-phuong.nguyen@u-bordeaux.fr}}

\date{2019 / 2019} %
\authorrunning{Phuong et al.} %
\titlerunning{Gas properties around GG Tau A}

\abstract
{GG Tau A is the prototype of a young triple T Tauri star that is surrounded 
by a massive and extended Keplerian outer disk. The central cavity is 
not devoid of gas and dust and at least GG Tau Aa exhibits its own
disk of gas and dust emitting at millimeter wavelengths. Its observed properties make this source an ideal 
laboratory for investigating planet formation in young multiple solar-type 
stars.}
{We used new ALMA  \tco~ and C$^{18}$O(3--2) observations 
obtained at high angular resolution ($\sim0.2''$) together with 
previous  CO(3--2) and (6--5) ALMA data and continuum maps at 1.3 
and 0.8\,mm in order to determine the gas properties (temperature, density, 
and kinematics) in the cavity and to a lesser extent in the outer disk.}
{By deprojecting, we studied the radial and azimuthal gas distribution 
and its kinematics. We also applied a new method to improve the 
deconvolution of the CO data and in particular better quantify the 
emission from gas inside the cavity. We perform local and nonlocal thermodynamic
equilibrium studies in order to determine the excitation conditions and 
relevant physical parameters inside the ring and in the central 
cavity.}
{Residual emission after removing a smooth-disk model indicates
unresolved structures at our angular resolution, probably in the
form of irregular rings or spirals. 
The outer disk is cold, with a temperature $< 20$\,K
beyond 250\,au that drops quickly ($\propto r^{-1}$). The kinematics
of the gas inside the cavity reveals infall motions at about 10\% of the
Keplerian speed. We derive the amount of gas in the cavity, and
find that the brightest clumps, which contain about 10\% of this mass,
have kinetic temperatures $40-80$\,K, CO column densities of
a few $10^{17}$\,cm$^{-2}$, and H$_2$ densities around $10^{7}$\,cm$^{-3}$.} 
{Although the gas in the cavity is only a small fraction of the disk
mass, the mass accretion rate throughout the cavity is comparable to
or higher than the stellar accretion rate. It is accordingly sufficient
to sustain the circumstellar disks on a long timescale.}


\maketitle{}
\section{Introduction}
\label{sec:intro}

In more than two decades of studying exoplanets, nearly 4,000 
exoplanets have been found. More than 10\% of these planets are 
detected in binary or higher hierarchical systems 
\citep{Roell+Neuh+Seifahrt_2012}. 
The general picture of planet formation is well agreed:
planets are formed within a few million years after the collapse phase
in a protoplanetary disk surrounding the protostar. 
However, the detailed formation conditions and mechanisms are still debated. 

\citet{Welsh+Orosz+Carter_2012} reported observations with the Kepler space 
telescope and revealed that planets can form and survive in binary systems, 
on circumbinary or circumstellar orbits.  The formation
conditions in these systems differ from those around single stars. 
Theoretical studies of disk evolution predict that the stars in a T Tauri binary of 
about 1\,Myr should be surrounded by two inner disks, that are located inside the 
Roche lobes, and by an outer ring or disk that is located outside the outer 
Lindblad resonances \citep{Artymowicz+Clarke+Lubow_1991}. For a binary 
system of low or moderate eccentricity, the stable zone is typically 
located beyond the 3:1 or 4:1 resonance  \citep{Artymowicz+Lubow_1994}. 
The residual gas and dust inside the nonstable zone inflows from the circumbinary to the
circumstellar disks through streamers  \citep{Artymowicz+Clarke+Lubow_1991}, feeding 
inner disks that otherwise would not survive.
The outer radii of these inner disks, as well as the inner radius of 
the circumbinary (outer) disk, are defined by tidal truncation. 
At (sub)millimeter wavelengths, circumbinary disks have been observed in many 
systems, and in some of these, such as L\,1551\,NE, UY Aur and GG Tau A 
\citep{Takakuwa+Saigo+Matsumoto_2014, Tang+Dutrey+Guilloteau_2014, 
Dutrey+DiFolco+Guilloteau_2014}, the circumstellar disk(s) is
also detected. Studying the gas and dust properties in these
environments is a necessary step for
understanding the formation of planets in the binary/multiple systems.  

The subject of this paper is a detailed study of gas and dust properties
of the GG Tau A system. 
GG Tau A, located in the Taurus-Auriga 
star-forming region consists of a single star GG Tau Aa and the 
close binary GG Tau Ab1/Ab2 with separations of 35\,au and 4.5\,au 
on the plane of the sky, respectively \citep{Dutrey+DiFolco+Beck_2016, 
DiFolco+Dutrey+LeBouquin_2014}. We here
ignore the binary nature of Ab, and only consider the 
GG Tau Aa/Ab binary, unless explicitly noted otherwise.
Although the GAIA results suggest a value of 150\,pc \citep{GAIA_2016, GAIA_2018}, 
we use a distance of 140\,pc for comparison with previous works. 

The triple star is surrounded by a Keplerian disk that 
is tidally truncated inside at $\sim$180\,au. The disk is comprised of a 
dense gas and dust ring that extends from about 180\,au up to 260\,au and 
an outer disk that extends out to $\sim$800\,au 
\citep{Dutrey+Guilloteau+Simon_1994}. The disk is inclined
at about $35^\circ$, with a rotation axis at PA\,$7^\circ$. The
Northern side is closest to us  
\citep{Guilloteau+Dutrey+Simon_1999}. The disk is one of the most 
massive in the Taurus region,  
$\sim$0.15\,M$_{\odot}$. This is $\sim$10\% of the total mass of the stars. A 
10\% mass ratio should lead to a small deviation (about 5\%) from the 
Keplerian law \citep{Hure+2011} outside the ring.

The outer disk is relatively cold and has dust and gas (derived from 
$^{13}$CO analysis) temperatures of 14\,K and 20\,K at 200\,au, respectively 
\citep{Dutrey+DiFolco+Guilloteau_2014, Guilloteau+Dutrey+Simon_1999}. 
More information about the triple system can be found in 
\citet{Dutrey+DiFolco+Beck_2016} and the references therein.
 
We here present a study of GG Tau A using sub-millimeter 
observations carried out with  ALMA. The main goal 
is to derive the properties of the gas cavity 
(density, temperature, and kinematics). For this purpose, we use 
a simple model for the gas and dust outer disk that allows us to better 
retrieve the amount of gas inside 
the cavity and simultaneously provides interesting information about the outer gas disk. 
The paper is organized as follows. Section \ref{sec:obs} describes the 
observations and data reduction. The observation results are presented 
in Section \ref{sec:res}, and the radiative transfer modeling of the 
outer disk is presented in  Section \ref{sec:disk}. The 
properties of the cavity (excitation conditions, mass, and dynamics) are 
derived in Section \ref{sec:cavity}. The gas and dust properties in 
the circumbinary disk and inside the tidal cavity are discussed in 
Section \ref{sec:disc}. Section \ref{sec:sum} summarizes the main 
results.  
\section{Observations and data reduction}
\label{sec:obs}
Table \ref{tab:obs1} lists the observational parameters of our
data sets: the spectral sampling, angular resolution and brightness
sensitivity for all
observed molecular lines. 
\begin{table*}[!ht] 
\centering
\caption{List of observations}
\label{tab:obs1} 
\begin{tabular}{ccccccc l }
\hline
{\bf{Spectral line}}& {\bf{Frequency}} & {\bf{Energy level}}& {\bf{Channel spacing}} & {\bf{Beamsize}} & {\bf{Noise}}\\
& {\bf{(GHz)}} & {\bf{(K)}}& {\bf{(km/s)}}  & & {\bf{(K)}} & ALMA project \\
\hline \hline
$^{12}$CO(6--5)& 691.473 & 33.2 & 0.106 & $0.35''\times0.31''$, PA=$104^\circ$ & 1.8 & 2011.0.00059.S\\
$^{12}$CO(3--2)& 345.795 & 16.6 & 0.106 & $0.34''\times0.28''$, PA=$ 91^\circ$ & 0.7 & 2012.1.00129.S \\
$^{13}$CO(3--2) & 330.588 & 15.9 & 0.110 & $0.22''\times0.16''$, PA=$ 15^\circ$ & 0.7 & 2012.1.00129.S, 2015.1.00224.S \\
C$^{18}$O(3--2)& 329.330 & 15.8 & 0.110 & $0.19''\times0.14''$, PA=$ 19^\circ$ & 1.9 & 2015.1.00224.S\\
Continuum  & 330.15 & -- & -- & $0.19''\times0.14''$, PA=$ 10^\circ$ &  0.03 & 2015.1.00224.S\\
\hline 
\end{tabular}
\tablefoot{Col.3 gives the upper state Energy level.}
\end{table*}

GG Tau A was observed with ALMA Band 9 during Cycle 0 
(2011.0.00059.S) and Band 7 during Cycle 1 (2012.1.00129.S) 
and Cycle 3 (2015.1.00224.S). Anne Dutrey is the PI of the three projects. 
\vspace{-0.4cm}
\paragraph{{\bf{Cycle 0}}}
Observations were made on 2012 August 13. The 
spectral windows covered the $^{12}$CO(6--5) line \citep[see][for 
details of the data reduction]{Dutrey+DiFolco+Guilloteau_2014}. These 
data were processed here with a restoring beam of 
\mbox{$0.35''\times0.31''$ at PA=$104^\circ$}.
\vspace{-0.4cm}
\paragraph{{\bf{Cycle 1}}}
Observations were taken on 2013 November 18 and 
19. The spectral windows covered the lines of 
$^{12}$CO(3--2) and $^{13}$CO(3--2) at high spectral resolution 
($0.11\,\kms$).  
Details of data reduction are given in 
\citet{Tang+Dutrey+Guilloteau+etal_2016}. The $^{12}$CO(3--2) images 
were obtained here with a restoring beam of $0.34''\times0.28''$ at 
PA=$-89^\circ$, and the $^{13}$CO(3--2) 
data were  
merged with new data acquired in Cycle 3.
\vspace{-0.4cm}
\paragraph{{\bf{Cycle 3}}}
Observations were carried on 2016 September 25 and 30
 with 37 and 38 useful antennas in 
configuration C40-6. The projected baselines range from 16\,m to 
3049\,m, and the total time on source is 1.4\,hours. The spectral setup 
covered the lines of $^{13}$CO(3--2) and C$^{18}$O(3--2)
at 330.588 and 329.330\,GHz in two windows, each covering 58.89\,MHz bandwidth with a high spectral 
resolution of 141\,kHz ($\sim$0.11\,$\kms$). 
The continuum was observed in two separate windows, one centered 
at 330.15\,GHz with 1875\,MHz bandwidth and 
the other centered at 342.00\,GHz with 117\,MHz bandwidth. 

Data were calibrated with the 
CASA\footnote{https://casa.nrao.edu/} software (version 4.7.0). 
We used the quasar J0510+1800 for phase and bandpass calibration. 
The absolute amplitude calibration was made using J0522-3627 
(flux $\sim$3.84\,Jy at 343.5\,GHz at the time of observations). The calibrated data were regrided
in velocity to the local standard of rest (LSR) frame using the task cvel, and were exported 
through UVFITS format to the GILDAS\footnote{https://www.iram.fr/IRAMFR/GILDAS/} 
package for further data processing. 

 GG Tau has significant proper motions:
\citet{Ducourant+Teixeira_2005} cited $[17,-19]$ mas per year, while 
\citet{Frink+Roser+Neuh_1997} reported $[11,-28]$ mas per year. These measurements
are affected by the multiplicity of the star, however. To realign our
observations, we assumed that the continuum ring is centered on 
the center of mass of the system which we took as the center
of coordinates for our images. We fit the continuum emission with the 
sum of a circular Gaussian (for the circumstellar disk around Aa) 
and an elliptical ring (for the dust ring) in the $uv$ plane 
\citep{Guilloteau+Dutrey+Simon_1999, Pietu+Gueth+HilyBlant_2011}.
The apparent motion of the ring gives a proper motion of $[9,-23]$ mas per year,
which we applied to all our data set. 
With these proper motions, the origin of the coordinates is at RA=4h\,32m\,30.3s and 
DEC=$17^{\circ}\,31'\,40''$ at epoch 2000.  

The imaging and deconvolution was made with natural weighting, 
and the images were cleaned down to the rms noise level with the $hogbom$ algorithm 
Channel maps of the observed
lines are presented in Appendix \ref{app:chan}, Figs.\ref{fig:13co32new}--\ref{fig:c18o32new}. 
In these data, the maximum recoverable scale is about 7$''$ (1000 au at the source distance).
The images were not corrected for primary beam attenuation: the beam size of 17$''$ yields
attenuation of 12\% at 500 au and 4\% at 300 au.
\section{Results}
\label{sec:res}
\subsection{Continuum emission}
\label{sub:cont}

\begin{figure}[h] 
  \centering
  \includegraphics[width=0.9\linewidth]{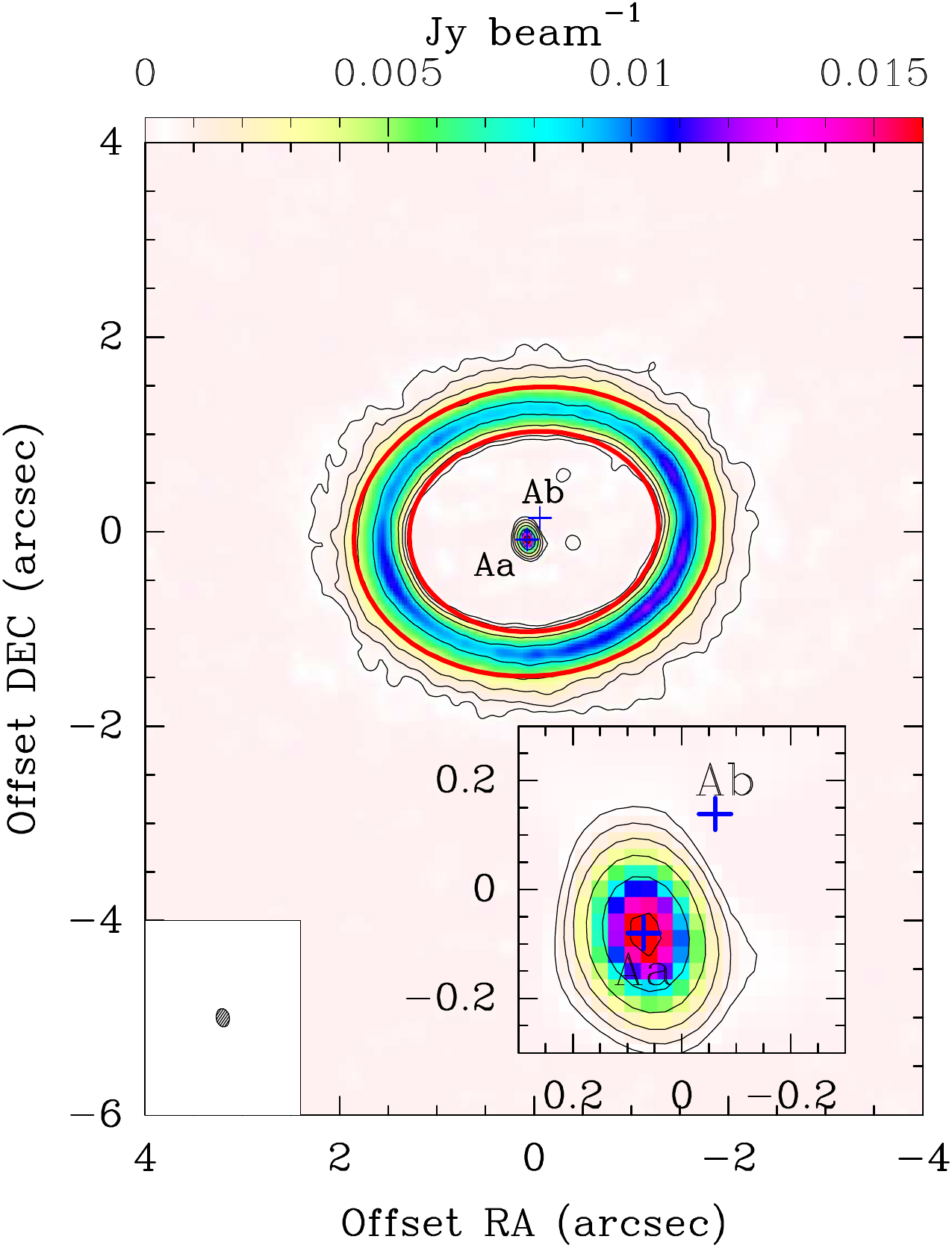}
  \caption{Continuum image at 330 GHz. The beam size of 
  \mbox{$0.19''\times0.14'',\rm{PA}=190^\circ$} is indicated in the 
  lower left corner. The contour levels are 0.5, 1, 2, 4, 8 and 16 
  mJy\,beam$^{-1}$. The noise level is 0.06\,mJy\,beam$^{-1}$. The red 
  ellipses indicate the inner and outer edges of the dense dust ring at 
 180\,au and 260\,au, respectively \citep[e.g.,][]{Guilloteau+Dutrey+Simon_1999}. The positions of the Aa star and Ab close
  binary are indicate by the crosses. The inset shows an enlarged
  view of the Aa and Ab surroundings.}
  \label{fig:cont}
\end{figure}

Figure \ref{fig:cont} shows the continuum emission at 330\,GHz 
derived from the Cycle 3 data. It reveals emission from the Aa 
disk and the ring structure detected in previous observations 
\citep[see][and references therein]{Dutrey+DiFolco+Beck_2016}, but the 
ring is now clearly resolved. The ring is not azimuthally symmetric 
(after taking the limb-brightening effect along the major 
axis into account), but displays a $\sim$15-20\% stronger emission at 
$\rm{PA}\approx$$240^\circ-260^\circ$. The outer edge is clearly 
shallower than the steep inner edge, confirming that some dust remains 
beyond the $\sim$260\,au outer edge of the ring, as initially 
mentioned by \citet{Dutrey+Guilloteau+Simon_1994}. 
As in previous studies \citep[e.g.,][]{Guilloteau+Dutrey+Simon_1999, 
Pietu+Gueth+HilyBlant_2011}, compact, unresolved emission is detected 
in the direction of the single star Aa, but no emission originates from 
the Ab close binary system.

A detailed study of the continuum emission is beyond the
scope of this paper and is deferred to subsequent work.

\begin{figure*}[ht!]  
  \centering
  \includegraphics[width=0.7\linewidth,trim=0cm -0.3cm 0cm 0cm]{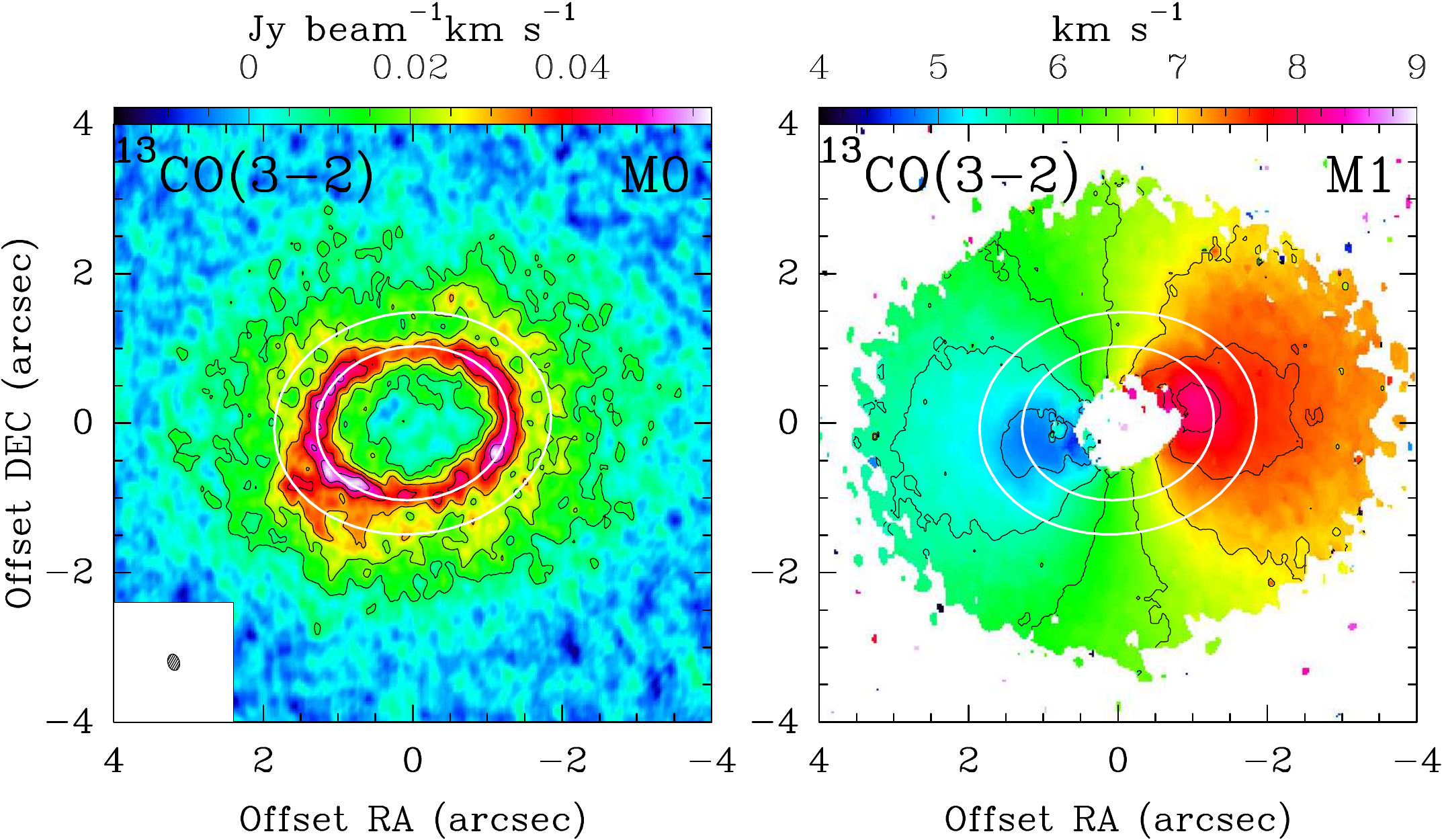}
  \includegraphics[width=0.7\linewidth, trim=0cm 0cm 0cm -0.3cm]{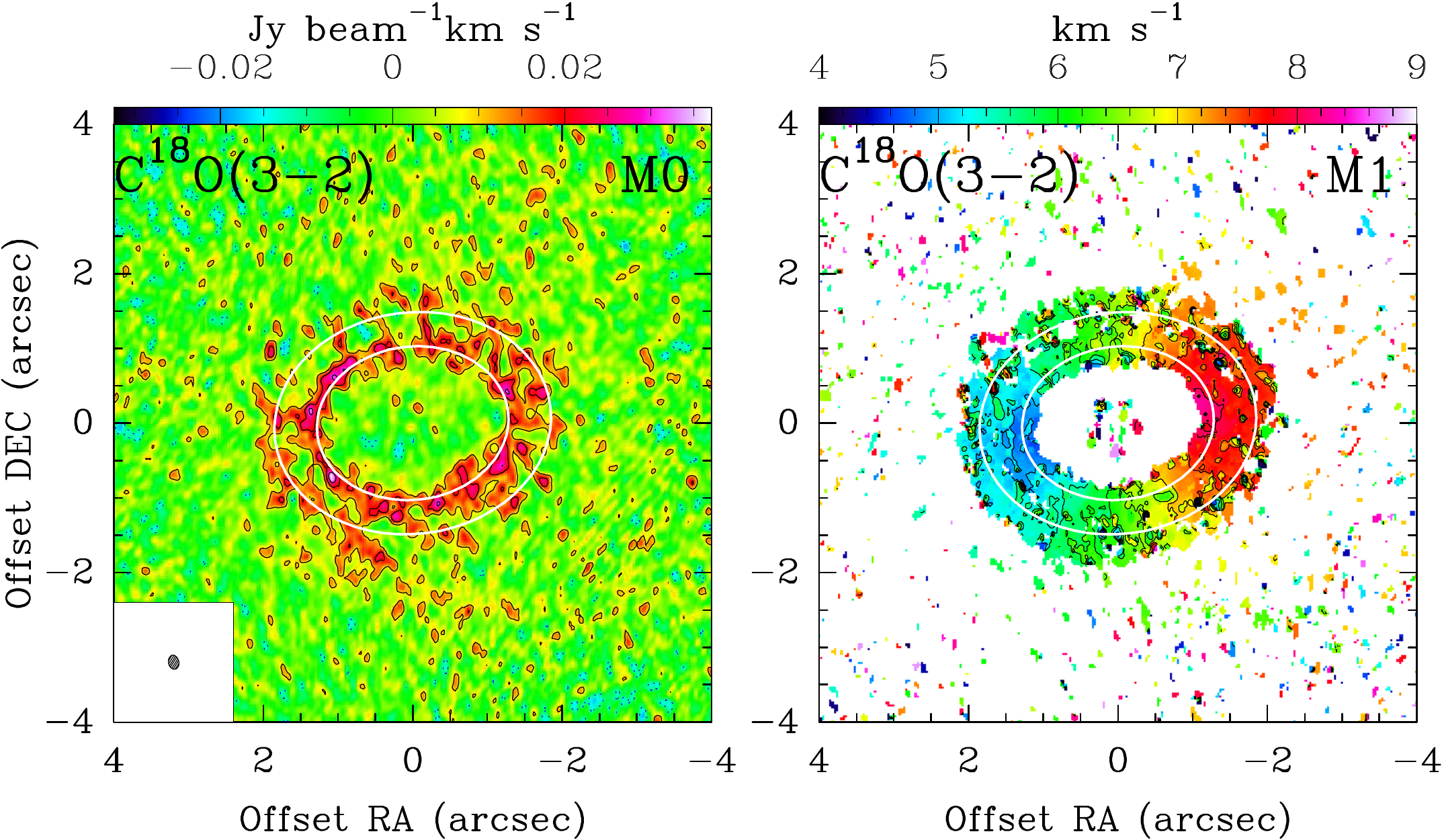}
  \caption{{\it{Upper:}} $^{13}$CO(3--2) integrated intensity map 
  (left, contour spacing 0.01 Jy/beam\,km\,s$^{-1}$ (3 $\sigma$) with zero level
  omitted) and velocity map (right). {\it{Lower:}} C$^{18}$O(3--2) 
  integrated intensity map (left, contour spacing 0.01 Jy/beam\,km\,s$^{-1}$, 2 $\sigma$)
  and velocity map (right).The beams 
  are indicated in lower left corner in each intensity map. The contour level
  spacing in the velocity maps is $0.5\kms$. The continuum has been subtracted. The white ellipses indicate the
  inner and outer edges of the dust ring.}
   \label{fig:isotop} 
\end{figure*}

\subsection{Images of line emission}
\label{sub:line}

Figure \ref{fig:isotop} shows the integrated intensity and the velocity 
maps of $^{13}$CO(3--2) (left) and C$^{18}$O(3--2) (right).
In these figures, the continuum has been subtracted. The velocity fields 
suggest Keplerian rotation inside the disk. 

The $^{13}$CO(3--2) emission extends out to 550\,au,
and the C$^{18}$O(3--2) is mostly visible in the dense 
ring. 

\begin{figure*}[ht] 
  \centering
  \includegraphics[height=7.20cm]{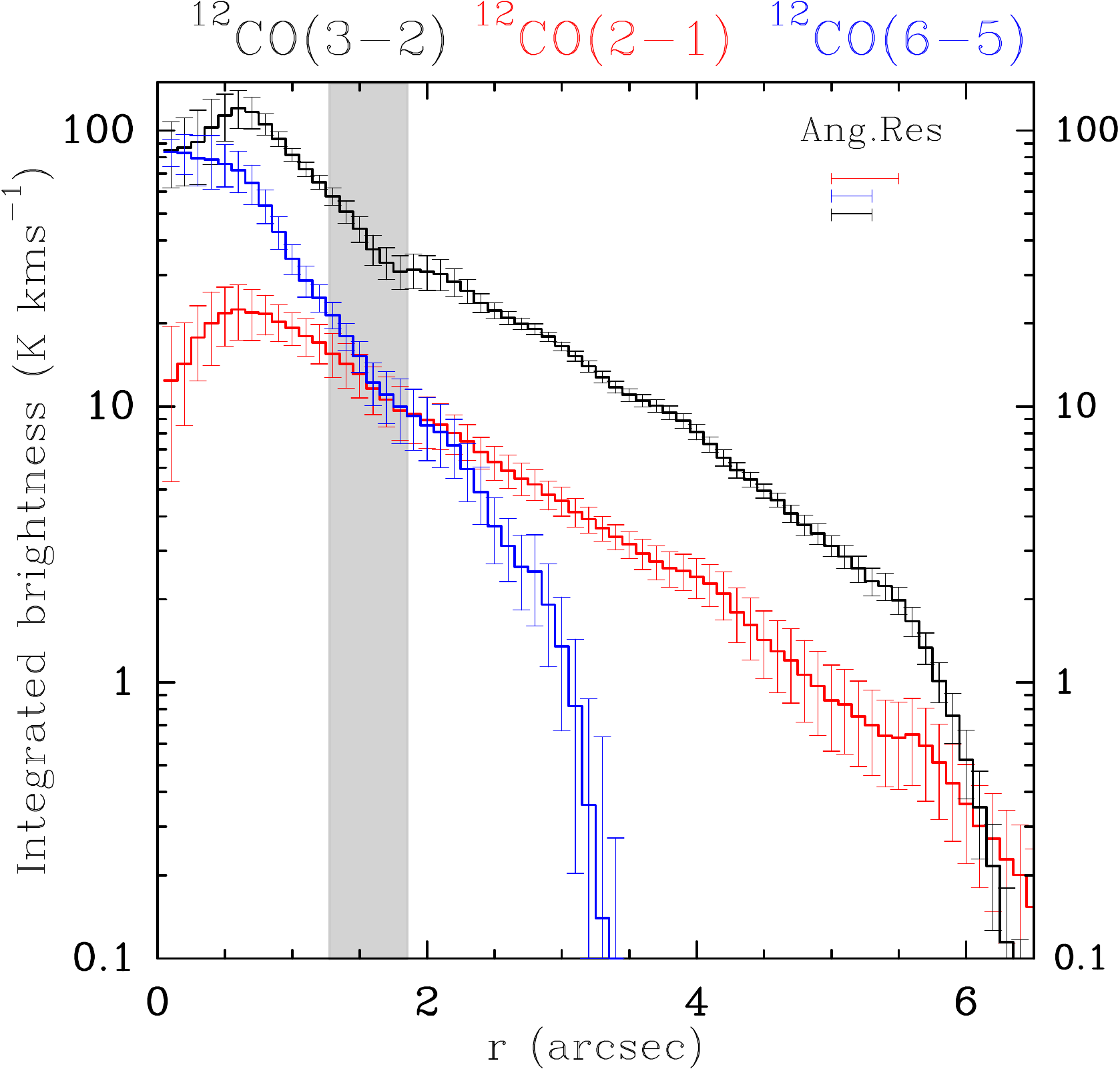}
  \hspace{0.15cm}
  \includegraphics[height=7.20cm]{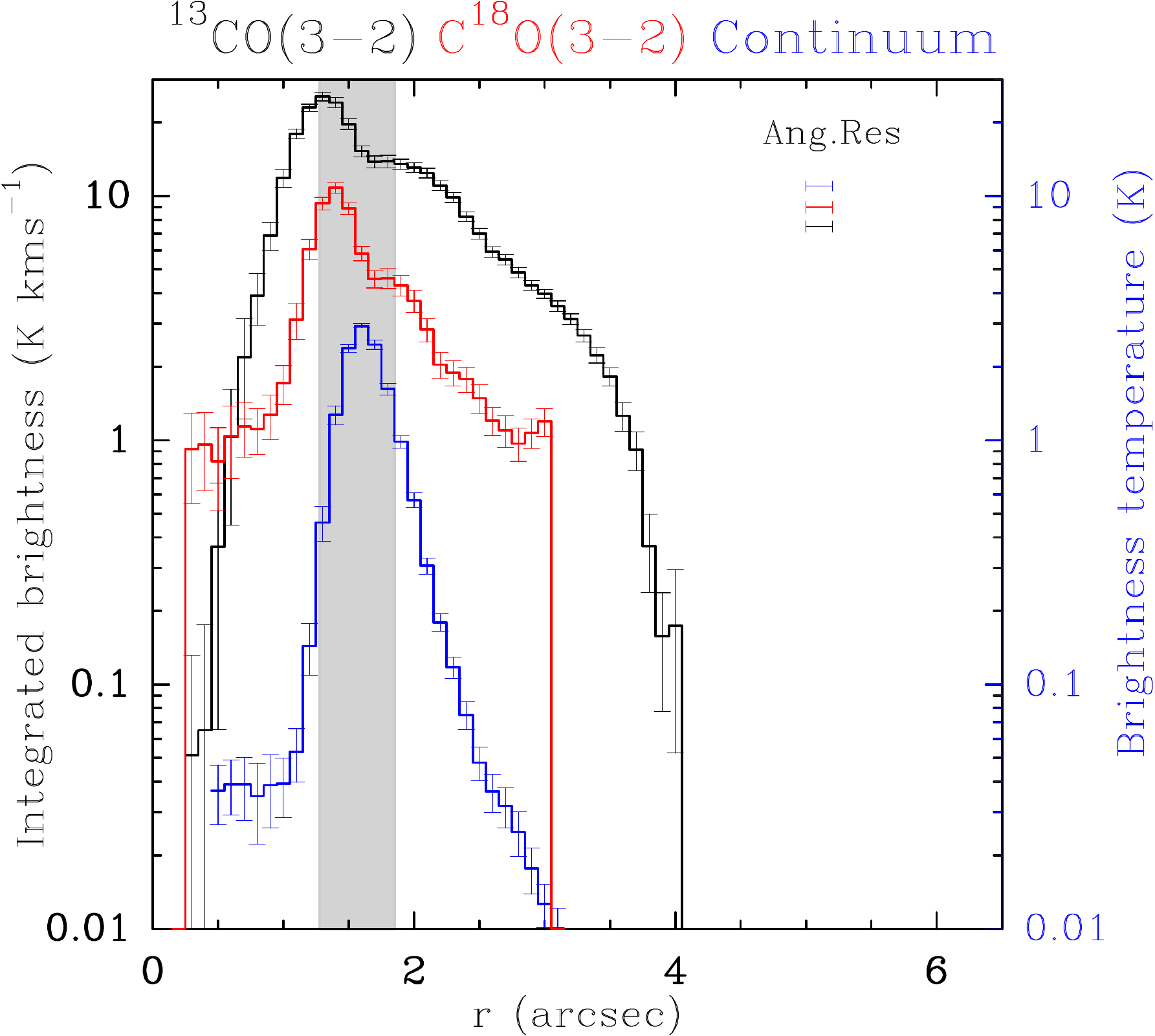} 
  \\
  \vspace{0.25cm}
  \includegraphics[height=7.20cm]{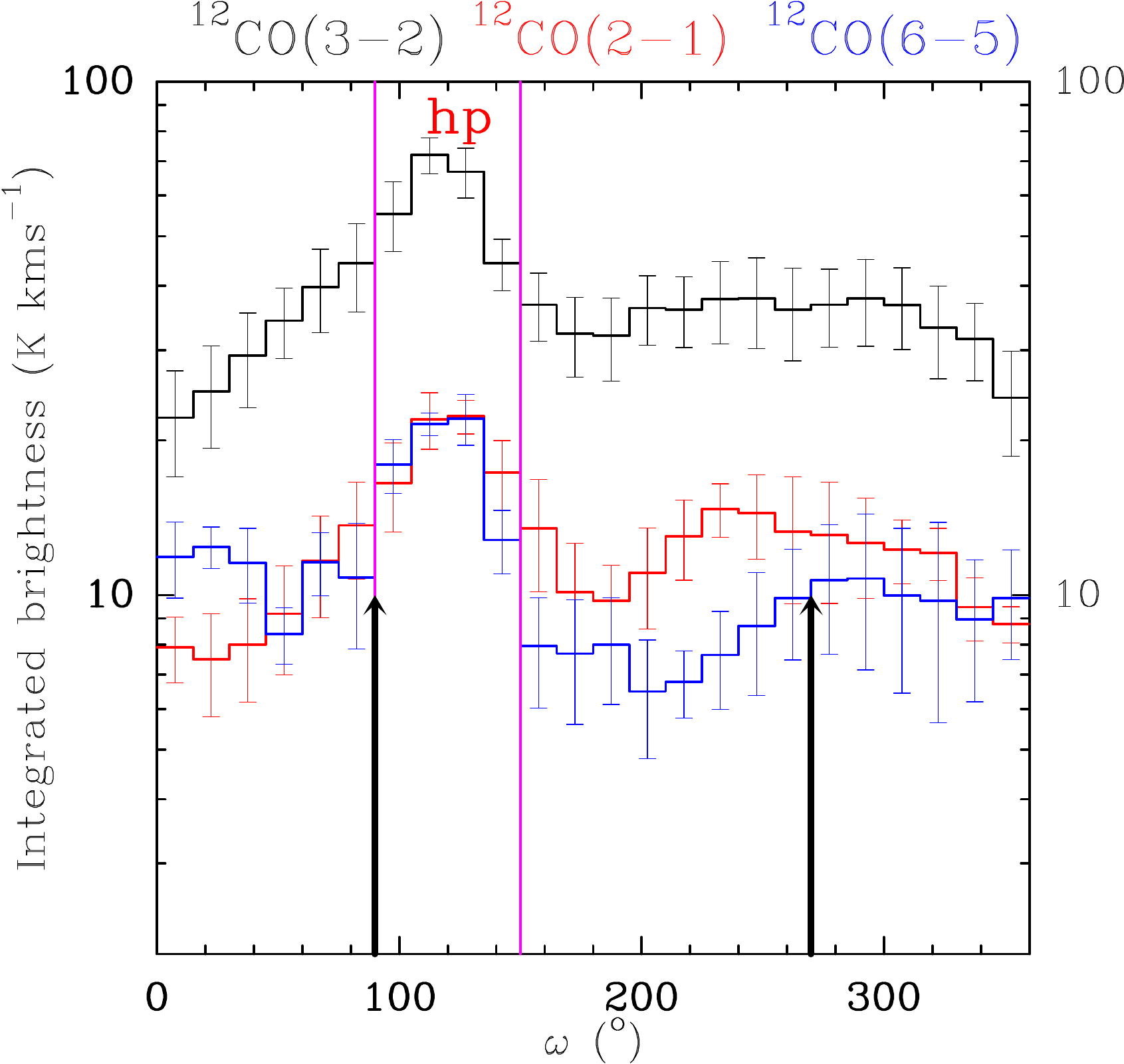}
  \hspace{0.15cm}
  \includegraphics[height=7.20cm]{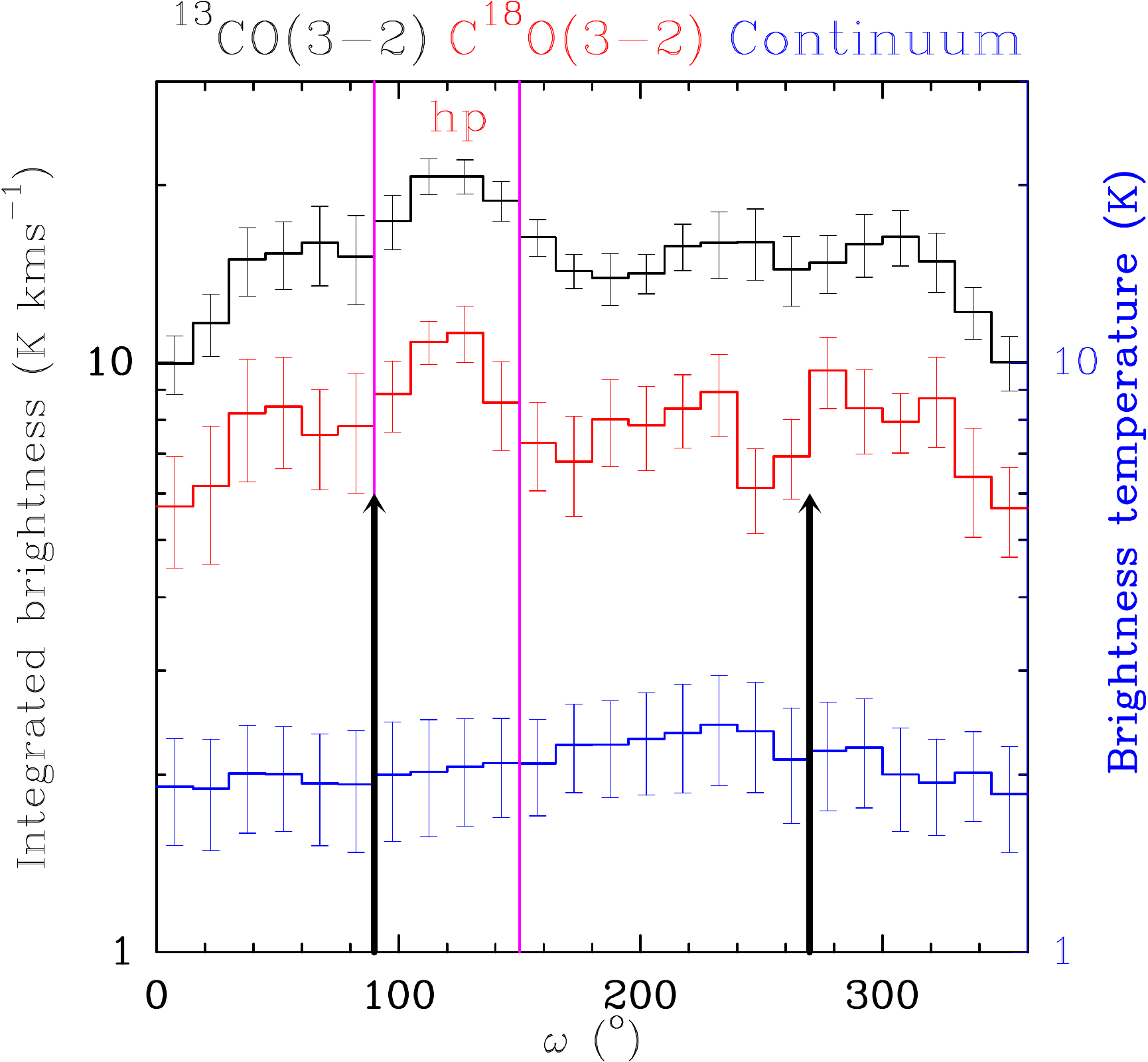}

\caption{{\it{Upper panels:}} Radial dependence of the integrated 
brightness temperature (for lines) and brightness temperature (for 
continuum emission) in the disk plane. The horizontal sticks indicate 
the angular resolutions. {\it{Lower panels:}} Azimuthal dependence of the 
same quantities averaged over the ring of $1.2''<r<2.0''$. 
The left panels show the plot of the emission from
the J=6--5, 3--2 and 2--1 lines of $^{12}$CO. For a description
of the corresponding data, see \citet{Tang+Dutrey+Guilloteau+etal_2016}
 for the J=3--2 line
and  \citet{Dutrey+DiFolco+Guilloteau_2014}
for the others.
 The right panels show the less 
abundant CO isotopologs (J=3--2) emissions. The gray region 
delineates the dust ring in the upper panels. In the lower panels, 
black arrows show the location of the limb-brightening peaks and the 
magenta lines show the hot-spot location.}
   \label{fig:profile}
\end{figure*}

\subsubsection{Intensity variations}
Figure \ref{fig:profile} (upper panels) shows the radial profiles of 
the integrated brightness of the lines and of the peak brightness of 
the continuum emission, averaged over the entire azimuthal direction, 
after deprojection to the disk plane.
The deprojection was made assuming a position angle of the 
minor disk axis of 7$^\circ$ and an inclination of 35$^\circ$ 
\citep{Dutrey+DiFolco+Guilloteau_2014, Phuong+Diep+Dutrey_2018}. 
See also \citet{Phuong+Diep+Dutrey_2018} for a detailed description of the
deprojection.

The $^{12}$C$^{16}$O emission covers a broad region around the central 
binary, $r\apprle6''$ (800\,au), peaking at the center. Some of 
the differences between the three transitions of CO may result 
from calibration effects and different $uv$ coverage. 
In particular, short spacings are missing in the CO(6--5) 
transition data because of its high frequency, making it 
less sensitive to extended structures.

Figure \ref{fig:profile} (lower panels) displays the azimuthal 
dependence (in the disk plane) of the peak brightness and 
velocity-integrated brightness in the ring ($1.2''\le r \le2''$) for CO, 
$^{13}$CO, C$^{18}$O and the 0.85\,mm continuum emissions. The azimuth $\omega$ in 
the disk midplane is measured counterclockwise from the minor axis, 
with 0 in the north direction. The significant enhancement in the 
southeastern quadrant  for $^{12}$C$^{16}$O corresponds to the 
hot spot discovered by 
\citet{Dutrey+DiFolco+Guilloteau_2014}, which may reveal a possible 
planet in formation (labelled \textquotedblleft 
hp\textquotedblright\,for hypothetical 
planet in the figure). It is far fainter in 
the other CO isotopologs.

\subsubsection{CO gas kinematics in the outer disk}
\label{sub:disk:kine}

\begin{figure}[h!] 
  \centering
    \includegraphics[width=0.9\linewidth]{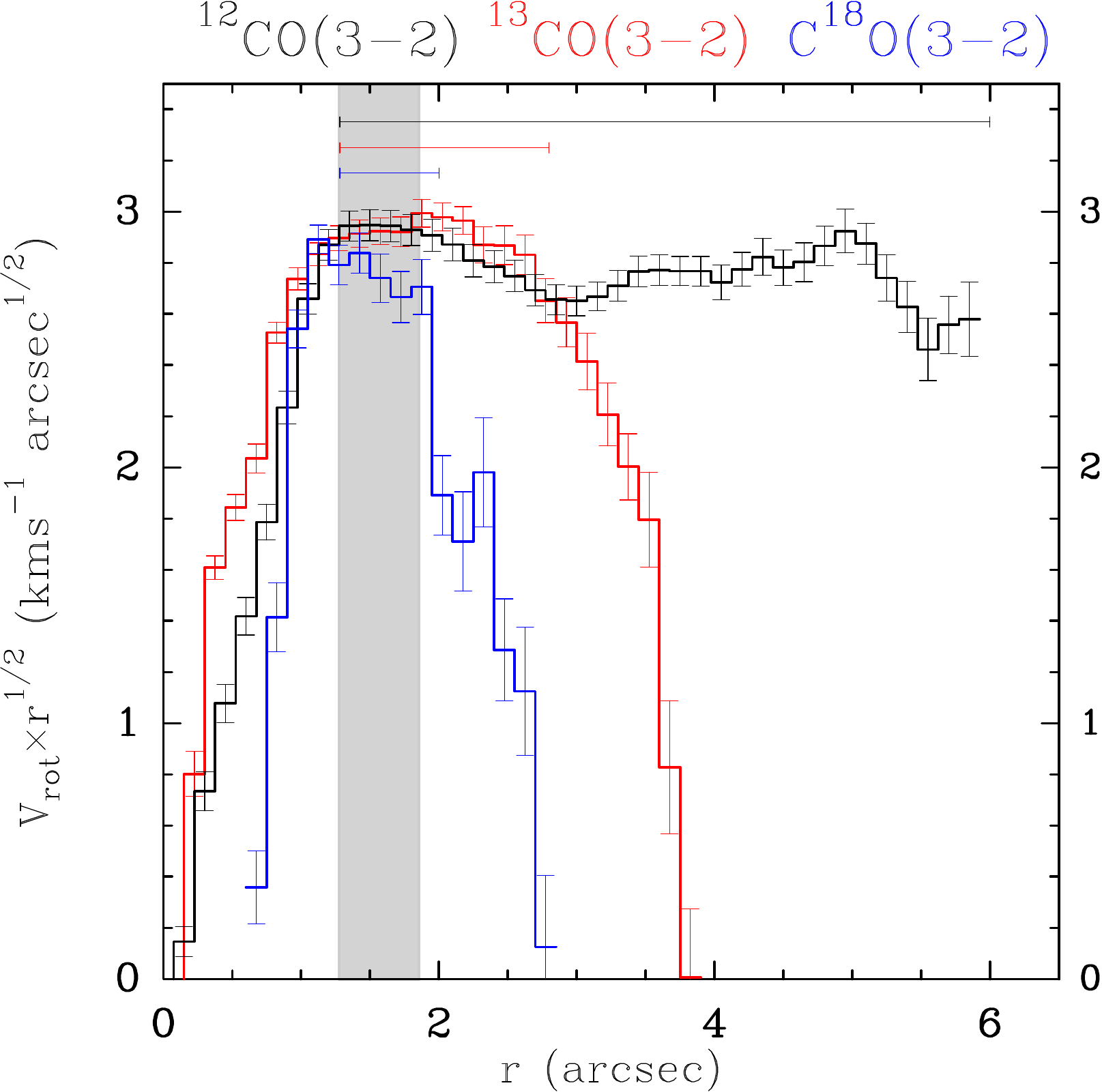}
  \caption{Dependence on $r$ of $V_{rot}\times\,r^{1/2}$ (weighted by 
  the brightness and averaged in bins of $0.15''$) of 
  $^{12}$CO(3--2) (black), $^{13}$CO(3--2) (red) and C$^{18}$O(3--2) 
  (blue) emissions. The horizontal bars indicate the radius
  range over which the mean value is computed for each transition. The gray shaded
  area indicates the boundaries of the dense dust ring.}
   \label{fig:vrot}
\end{figure}

For a thin disk, the line-of-sight velocity is given by
\mbox{$V_z=\sin i(V_{rot}\sin\omega+V_{fall}\cos\omega)$} 
where $i$ is the disk 
inclination, $V_{rot}$ the rotation velocity and $V_{fall}$ the infall 
velocity, and $\omega$ is the azimuth in the disk plane. 
We can neglect the infall motions, because the studies of \citet{Phuong+Diep+Dutrey_2018}  at 
angular resolution of $\sim0.35''$ have placed an upper limit of 
9\% on it with respect to the rotation. We used the intensity-weighted
images of the line-of-sight velocity $V_z$ shown in Fig.\ref{fig:isotop} 
and for each pixel calculate $V_z(\sin i\sin\omega)^{-1}$.
We then azimuthally averaged these values for all pixels at the
same radius (using a $0.15''$ radial binning)
such that $|\sin\omega|$ to $> 0.7$ (i.e., avoiding pixels around the
minor axis) to derive $V_{rot}$.

Figure \ref{fig:vrot} shows the dependence of 
$\langle V_{rot}\,(r/1'')^{1/2}\rangle$ on $r$, which would be constant for the 
three CO isotopologs if the rotation were Keplerian.
The overall agreement between the three isotopes is good, showing
that the outer ring and disk are in Keplerian rotation beyond
about 180\,au. A constant fit to 
these histograms gives $V_{rot} \approx 2.94\kms$, with a
standard deviation ($\sigma$) of $0.14\kms$ of the residuals from the mean for CO, 
$2.98\kms$ ($\sigma = 0.04\kms$) for $^{13}$CO 
and $2.81\kms$ ($\sigma = 0.07\kms$) for C$^{18}$O, at $\approx1.0''$
(the integration ranges are illustrated in Fig.\ref{fig:vrot}).

The formal errors on these mean values are two to three times smaller,
depending on the number of independent points, which is not a simple 
value given our averaging method. However, the CO data show
deviations from the mean that are not random, because they occur
on a radial scale of $\approx0.8''$, more than twice the resolution.
We therefore conservatively used the standard deviation as the error 
on the mean. Using the three independent measurements from CO, $^{13}$CO,
and C$^{18}$O, we derive a mean weighted value of $2.94\pm0.03\kms$ for the 
Keplerian rotation speed at $1''$, i.e. $3.48\pm0.04\kms$ at 100\,au, in 
agreement with previous, less precise determinations (e.g. 
\citet{Dutrey+DiFolco+Guilloteau_2014} found $3.4\pm0.1\kms$). 
When the uncertainty on the inclination, $\pm2^\circ$, is taken into
account, this 
corresponds to a total stellar mass of $1.36 \pm 0.07\msun$.

The apparent lower-than-Keplerian velocities for radii larger 
than $2.1''$ (for C$^{18}$O) or $3''$ (for $^{13}$CO) are due to the 
low level of emission in these transitions. Those at radii 
smaller than about $0.8''$ are discussed in 
Sec.\ref{sec:sub:cavity}.
\section{Disk modeling}
\label{sec:disk}

\subsection{Objectives}
Our main objective is to derive the properties of the gas
in the cavity. However, the overall source size, in CO and \tco, 
is comparable to or larger than the largest angular scale that can 
be recovered. This means that a simple method in which we would just use the globally deconvolved image
to identify the emission from the cavity might be seriously affected by
the missing extended flux.  Accordingly, we used a more elaborate approach in
which we fit a smooth-disk model to the emission beyond about 160 au.
The extended emission from this model is better recovered through its analytic
prescription. This outer disk model was later subtracted 
from the visibility data, so that the residuals have no structures at scales larger
than about 3.5$''$, that is, the cavity size. These residuals can then be
properly imaged and deconvolved without suffering from missing large-scale flux.

\subsection{Disk model}

We used the DiskFit tool \citep{Pietu+Dutrey+Guilloteau+etal_2007} to 
derive the bulk properties of the ring and outer disk. The disk model 
was that of a flared disk with piece-wise power laws for the 
temperatures and surface densities, and sharp inner and outer 
radii beyond which no molecule (or dust) exists. The temperature was 
vertically isothermal, and the vertical 
density profile was a Gaussian, $n(r,z) = n_0(r) \exp(-(z/H(r))^2)$,
with a scale height following a simple power 
law $H(r)=h_0(r/R_0)$.
For spectral lines, we assumed that the velocity field 
was Keplerian, $v(r) = V_0(r/R_0)^{-0.5}$, and used a constant local line
width $dV$. The lines are assumed to be at local thermal equilibrium (LTE): 
the derived temperatures thus 
indicate excitation temperatures. 
The emission from the disk was computed using a ray-tracing method.
The difference between the predicted model visibilities and the 
observed ones is minimized using a Levenberg--Marquardt method, 
and the error bars were derived from the covariance matrix.

A more detailed description of the DiskFit tool and our adopted
fitting method is given in Appendix \ref{app:diskfit}.
The fit parameters are summarized in Table \ref{tab:diskfit}.

\subsection{Continuum fit}

The CO emission being at least partially optically thick, we cannot 
simply separate the contribution of CO and continuum 
\citep{Weaver+Isella+Boehler_2018}. To determine the continuum 
properties, we fit the continuum using the broadband, line-free 
spectral window data. We followed the procedure described in 
\citet{Dutrey+DiFolco+Guilloteau_2014}, who derived dust 
properties using 1.3\,mm and 0.45\,mm continuum. We first subtracted a 
Gaussian source model of the emission from the circumstellar disk of 
Aa. The emission from the ring was then fit by a simple power-law 
distribution for the surface density and temperature, with 
sharp inner and outer edges (see also Appendix \ref{app:diskfit}), 
assuming a spatially constant dust absorption coefficient that scales 
with frequency $\nu$ as $\nu^\beta$. 

The goal of this continuum modeling is that residual emission
after model fitting becomes lower than the noise level in the
spectral line data, so that the continuum does not introduce
any significant bias in the combined fit for spectral lines described
in the next section. It is enough to adjust only the surface density for
this. We thus fixed geometrical parameters (center, position angle, and inclination, see Table \ref{tab:geometry}) 
and the inner and outer radii, the absorption coefficient, and the 
temperature profile, using the results of \citet{Dutrey+DiFolco+Guilloteau_2014}.
The results of the continuum fit are given in Table \ref{tab:cont}.
The residuals, such as those due to the 
$\simeq 20$\,\% azimuthal variations visible in Fig.\ref{fig:cont}, or
the shallow outer edge of the brightness distribution, are well
below 1 K in brightness.

\begin{table}[h!]   
\caption{System geometric and kinematic parameters}
\begin{center}
 \begin{tabular}{lcll}
\hline
Parameter &  Value &  & \\
\hline
$(x_0,y_0)$          &  (0,0)  & Center of dust ring  & fixed \\
$PA (^{\circ})$      &   $7$   & PA of disk rotation axis & fixed \\
$i (^{\circ}) $      &  $-35$  & Inclination & fixed \\
$V_\mathrm{LSR}$ ($\kms$)    &  6.40   & Systemic velocity & fixed \\
$V_0$ (km\,s$^{-1})$ & $3.37$  & Keplerian velocity at 100 au & fixed \\
$dV$ (km\,s$^{-1})$  & $0.3$   & Local line width & fixed \\
\hline
\end{tabular}
\end{center}
\label{tab:geometry} 
\vspace{-0.5cm}
\tablefoot{See Appendix B for a detailed description of the
handling of ``fixed'' values.}
\end{table}
\begin{table}[h!] 
\caption{Dust ring parameters}
\begin{center}
\begin{tabular}{llc}
\hline
 Parameter & Value/Law &  \\
\hline
Inner radius (au) & 193 &  fixed \\
Outer radius (au) & 285 &  fixed \\  
Abs. coefficient
$\kappa_{\nu}$ (cm$^{2}/g$) & $ 0.02 \times (\nu/230\mathrm{GHz})^{+1}$  & fixed \\    
Temperature (K) & $14  \times (r/200\mathrm{\,au})^{-1}$ & fixed  \\
Surface density (cm$^{-2}$) & 5.6 $10^{24} \times (r/200\mathrm{\,au})^{-1.4}$ & Fitted\\
\hline                           
\end{tabular}
\end{center}
\label{tab:cont} 
\vspace{-0.5cm}
\tablefoot{Fixed values are taken from \citet{Dutrey+DiFolco+Guilloteau_2014}.
See Appendix B for a detailed description of the
handling of ``fixed'' values.}
\end{table}
\subsection{CO isotopologs}
\label{sec:sub:coiso}

We analyzed the CO isotopolog data without removing the continuum.
The parameters labeled ``fixed'' in Tables \ref{tab:geometry} - \ref{tab:cont} were
used as fixed input parameters for our modeling.

While the outer disk is well represented by a Keplerian disk,
the  emission from the cavity does not follow such a simple
model. It contributes to a significant fraction of the total emission from CO, however.
Because the fit was made by minimizing in the visibility (Fourier) plane,
we cannot separate the cavity from the contributions of outer disk in this 
process.

To avoid biasing our results for the ring and disk 
parameters, we therefore adopted a specific strategy. We first subtracted the 
Clean components located inside the cavity (up to a radius of 
160\,au) from the original $uv$ tables (this also removes the 
continuum from Aa). The modified $uv$ tables were then analyzed using an 
innerly truncated Keplerian disk model, as described in detail
in Appendix \ref{app:diskfit}.

Because the radial profiles of the emission from CO and $^{13}$CO are not well 
represented by a power law 
\citep[see Fig.\ref{fig:profile}, and Fig.3 of][]{Tang+Dutrey+Guilloteau+etal_2016}, our disk model assumes 
piece-wise, continuous power laws (linear in log-log space) for the 
surface density and temperature. We used knots at 160, 200, 260, 300 
and 400\,au. 
The knots were chosen to reflect both the slope changes
in the radial profile of the line emissions and the sharp edges of the dust ring
based on our previous studies \citep{Dutrey+Guilloteau+Simon_1994, Guilloteau+Dutrey+Simon_1999},
 and to allow a good estimate for the properties of the bulk of the gas in 
the ring and outer disk using a minimum of knots.

The following strategy was adopted to fit in parallel the $\dco$ 
and $\tco$ data. In a first step, we determined the temperature by 
fitting  the $\dco$ line. The surface density of CO is not a critical 
value here: because CO  is largely optically thick, we just need to use a 
high enough CO surface density to ensure this. We then used this 
temperature profile to fit the $\tco$ data and determine the $\tco$ 
surface density, because this line is partially optically thin.  The 
derived surface density was then multiplied by the isotopic ratio 
$\dco/\tco$ \citep[70,][]{Milam+Savage+Brewster_2005} to specify the CO 
surface density to iterate on the temperature determination using the 
$\dco$ data. The process converges quickly (in two iterations).  

Our method makes the underlying assumption that the $\dco$ and $\tco$ 
layers are at the same temperature. This hypothesis is consistent with 
the results from \citet{Tang+Dutrey+Guilloteau+etal_2016}, who found 
that the vertical temperature gradient around $200-400$\, au needs to  
be small to reproduce the observed  $\dco/\tco$ line ratio. 

A fixed inner radius of 169\,au provided a good compromise to represent 
all molecular distributions. This radius is here only to obtain a good 
model for the ring and outer disk: it should not be overinterpreted as 
the physical edge of the cavity. We also independently determined the 
outer radii for each CO isotopolog, and verified the best-fit value 
for the inclination and systemic velocity. The small difference between 
our adopted Keplerian rotation law and the law suggested by the analysis 
in Sec.\ref{sub:disk:kine} has negligible effect on the fitted 
parameters. For the C$^{18}$O, we used only four knots to derive the surface 
density profile.

\begin{figure*} 
  \centering
  \includegraphics[width=5.8cm]{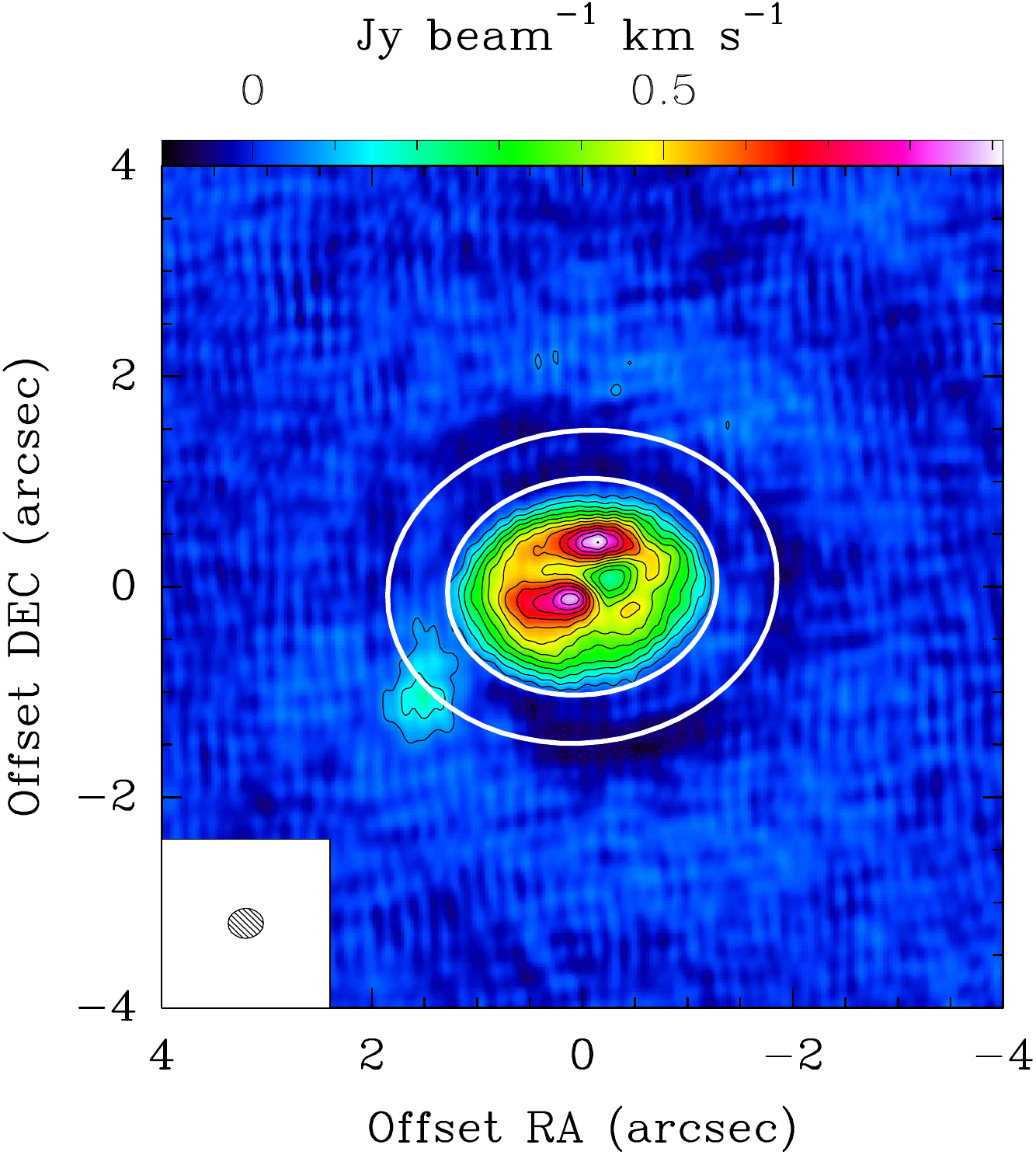}
  \hspace{0.3cm}
  \includegraphics[width=5.8cm]{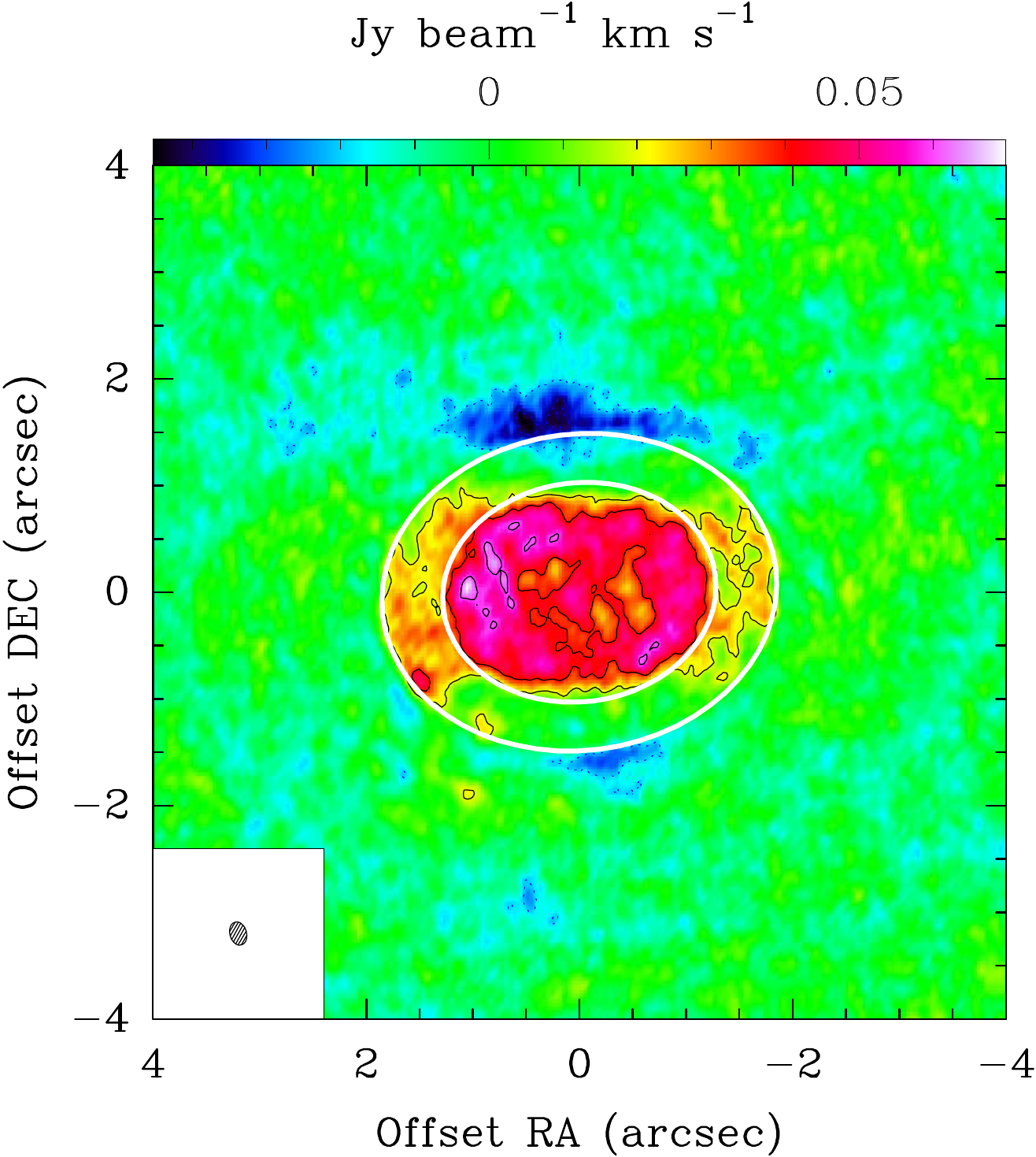}
  \hspace{0.3cm}
  \includegraphics[width=5.8cm]{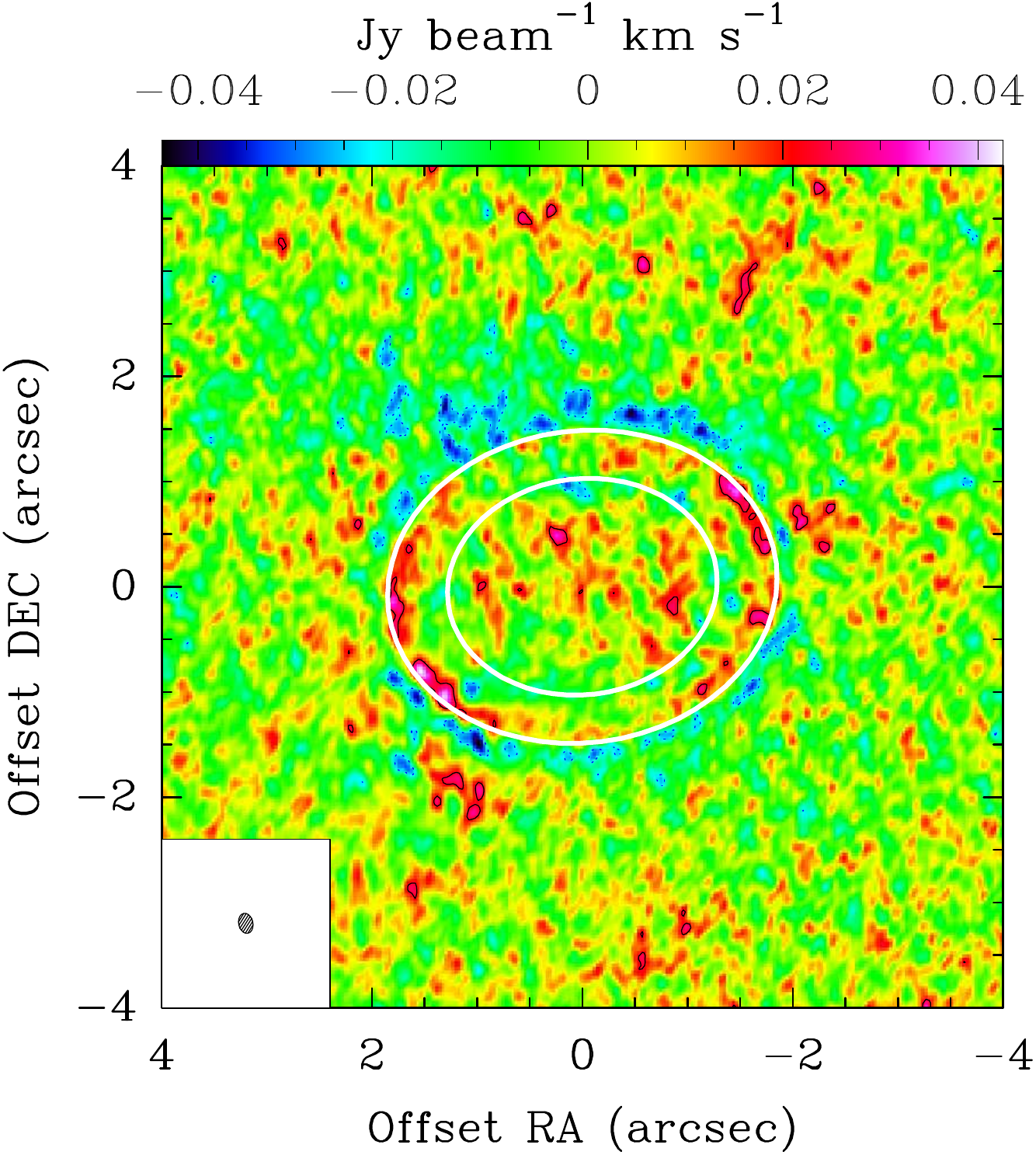}
\caption{Integrated intensity of (residual) emissions after subtracting 
the best ring+disk models. {\it{Left:}} $\dco$(3--2). The restoring beam of 
$0.34''\times0.28'', \rm{PA}=-89^\circ$ is indicated in the lower left 
corner. 
{\it{Middle:}} 
$\tco$(3--2). The restoring beam is \mbox{$0.22''\times0.16'', \rm{PA}=16^\circ$}.
{\it{Right:}} C$^{18}$O(3--2). The restoring beam is 
\mbox{$0.19''\times0.14'', \rm{PA}=-167^\circ$}.
The contour level is 0.07\,Jy\,beam$^{-1}$ km s$^{-1}$ (5$\sigma$) 
for $\dco$(3--2) map and $\sim$0.02\,Jy\,beam$^{-1}$ km s$^{-1}$ (3$\sigma$) 
for $\tco$(3--2) and $\cdo$(3--2) maps, the zero level is omitted.
The ellipses show the inner and outer edges of the dust ring at 180 and 260\,au.} 
   \label{fig:residual} 
\end{figure*}

\begin{figure}  
    \centering
     \includegraphics[width=0.88\linewidth]{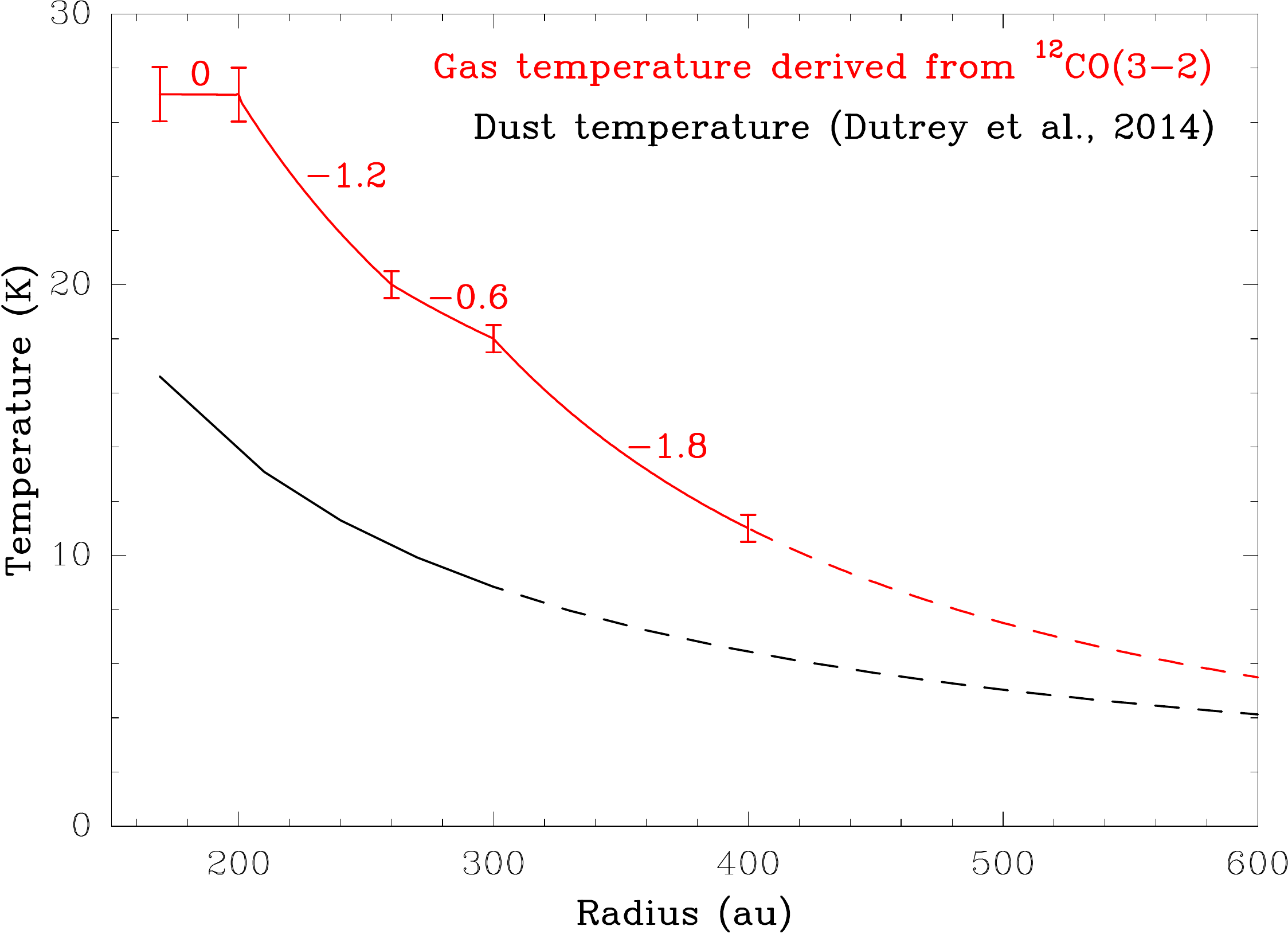} 
\caption{Radial dependence of CO gas (red) and dust (black) 
temperature. The gas temperature is derived from the $\dco$(3--2) 
analysis. Beyond 400\,au, CO temperature is extrapolated from the 
fitted law in the range of $300-400$\,au. The power-law exponents are 
indicated in each range. The uncertainty on the exponent is about 0.15 
for the best constrained values (-1.2 and -1.8), and 0.3 for the 
others. The primary beam attenuation introduces a bias of 
$8\%$ at 400 au. The dust temperature is taken from 
\citet{Dutrey+DiFolco+Guilloteau_2014}. Beyond a radius of 285 au, the 
dust temperature corresponds to an extrapolation.}
   \label{fig:cotemp}
\end{figure}

\begin{figure} 
    \centering
     \includegraphics[width=0.88\linewidth]{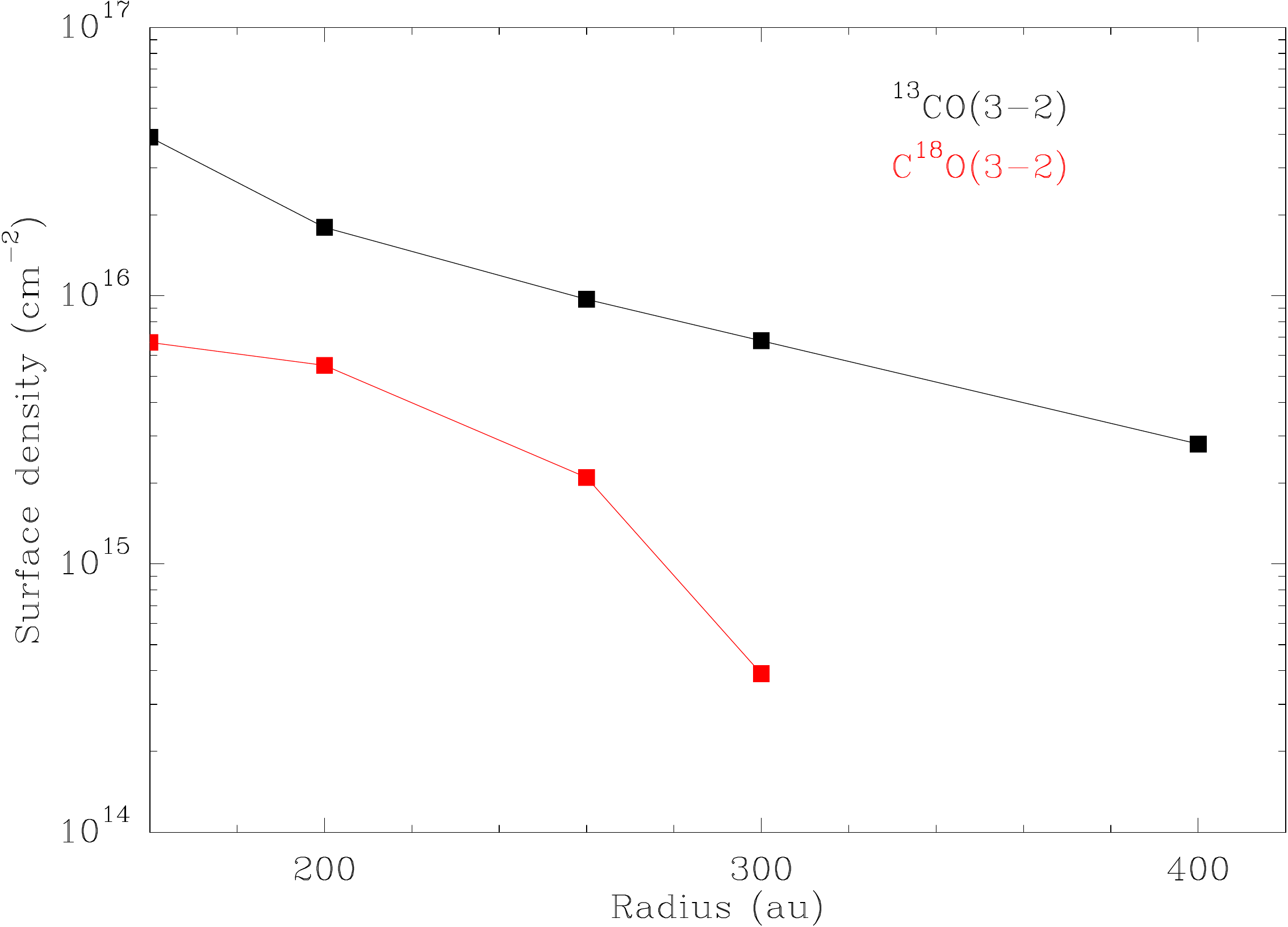}
\caption{Comparison of the surface densities from the LTE analysis.}
   \label{fig:cdplot}
\end{figure}

With this process, we find a reasonable model of the ring and outer 
disk in all CO isotopologs. Figure \ref{fig:residual}  
shows the residuals from the original $uv$ data after removal 
of the best-fit outer disk models and of Aa continuum source. As 
expected, most of the left-over emission is coming from the cavity, but 
some azimuthal asymmetries are still visible in the dense ring and 
beyond  (e.g. the hot spot and low level extended emission). The 
best fit results and formal errors are summarized in Table 
\ref{tab:co-final}, Figure \ref{fig:cotemp}, and Figure \ref{fig:cdplot}. Significant deviations from the best fit 
model do exist (e.g., azimuthal variations), therefore the results must be 
interpreted with caution. The formal errors underestimate the 
uncertainties on the physical parameters. We therefore also quote a 
confidence interval for the temperatures in Table \ref{tab:co-final}, 
based on the dispersion of values found during our minimization 
studies: surface densities are typically uncertain by $20-30$\%,
but the steep decrease in temperature between 200 
and 300\,au, and then to 400\,au and beyond is a robust result.
The surface density profile around $180-200$\,au is poorly constrained, 
because emission inside 160\,au was removed, and the 
angular resolution at this level is insufficient. However, the variations in the fitted 
surface densities between 169\,au (the inner truncation radius) and 
180\,au suggest a very dense inner edge.

In spite of its limitations, our approach leads to some conclusions.
Beyond a radius of about 200\,au, the CO gas is cold and temperatures 
drop from about 27\,K at 180\,au to 11\,K 
at 400\,au (see Fig.\ref{fig:cotemp}) (or about 13\,K after correction
for primary beam attenuation).

We note that the scale height of $24$\,au at 200\,au that was 
found to represent well the CO isotopolog emissions (see Appendix 
\ref{app:diskfit}) corresponds to a kinetic temperature of 15\,K under 
the hyothesis of hydrostatic equilibrium. This agrees reasonably well 
with the dust temperature derived by 
\citet{Dutrey+DiFolco+Guilloteau_2014}.

The outer radius of the  disk is $370$\,au in $\cdo$,
about 550\,au in $\tco$ and greater than 600\,au in $\dco$. 
The last two radii are less well constrained than that of $\cdo$ because
the temperature drops steeply with radius.  

\begin{table}[h!] 
\caption{Temperature derived from $^{12}$CO(3--2) and surface density from $^{13}$CO(3--2) and $\cdo$(3--2)} 
\begin{center}
\setlength{\tabcolsep}{4.5pt} 
\begin{tabular}{ccccccc}
\hline
(1) & (2) & (3) & (4) & (5) & (6) & (7) )\\
 r  &  T$_k$ & & & $^{13}$CO & C$^{18}$O & Ratio \\
(au) & (K)  & (K) & (K) &  \multicolumn{2}{c}{$10^{15}$\,cm$^{-2}$} & \\
\hline
160 & $27.2$ & $0.17$ & [26,28] & $ 39\pm2$  & $6.7\pm0.6 $ & $5.8\pm0.8$\\
200 & $27.4$ & $0.11$ & [25,28] & $ 18\pm1$  & $5.5\pm0.4$ & $3.3\pm0.4$ \\ 
260 & $19.7$ & $0.07$ & [19,21] & $9.7\pm0.3$ & $2.1\pm0.1$  & $4.6\pm0.4$  \\
300 & $18.0$ & $0.03$ & [17,19] & $6.8\pm0.1$  & $0.39\pm0.02$& $17\pm1$ \\
400 & $10.7$ & $0.02$ & [10,12] & $2.8\pm0.03$ & -- & --\\
\hline                                    
\end{tabular}
\end{center}
\label{tab:co-final} 
\vspace{-0.5cm}
\tablefoot{Nominal model fit after removal of Clean Components for
$r<160$\,au.
(1) Knot radius. (2) Temperature derived from $\dco$(3--2)
and (3) its formal error from the fit and (4) estimated confidence interval
from the minimizations. These uncertainty estimates do not take into account
an additional 10$\%$ error on the absolute flux scale.
(5-6) molecular surface density, (7) $\tco$/$\cdo$ surface density ratio.}
\end{table}

\section{Analysis of the gas inside the cavity}
\label{sec:cavity}

With a good first-order model for the spectral line emission in 
the ring and outer disk (i.e. beyond 169\,au), we can now determine a better 
representation of the emission coming from the gas in the cavity. 
For this purpose, we subtracted our best ring+disk model (presented in Sec.\,\ref{sec:disk})
from the original visibilities. CLEANed images of this residual emission, which
mostly comes from the cavity, were produced for the three CO isotopologs  
($^{12}$CO, $^{13}$CO and C$^{18}$O J=3--2).  
Figure \ref{fig:residual} presents the residual maps obtained. 
In this section, we study the properties of
the gas inside the cavity using these residual maps. 

\subsection{Dynamics inside the cavity} 
\label{sec:dynamics}

\begin{figure} 
    \centering
     \includegraphics[width=\linewidth]{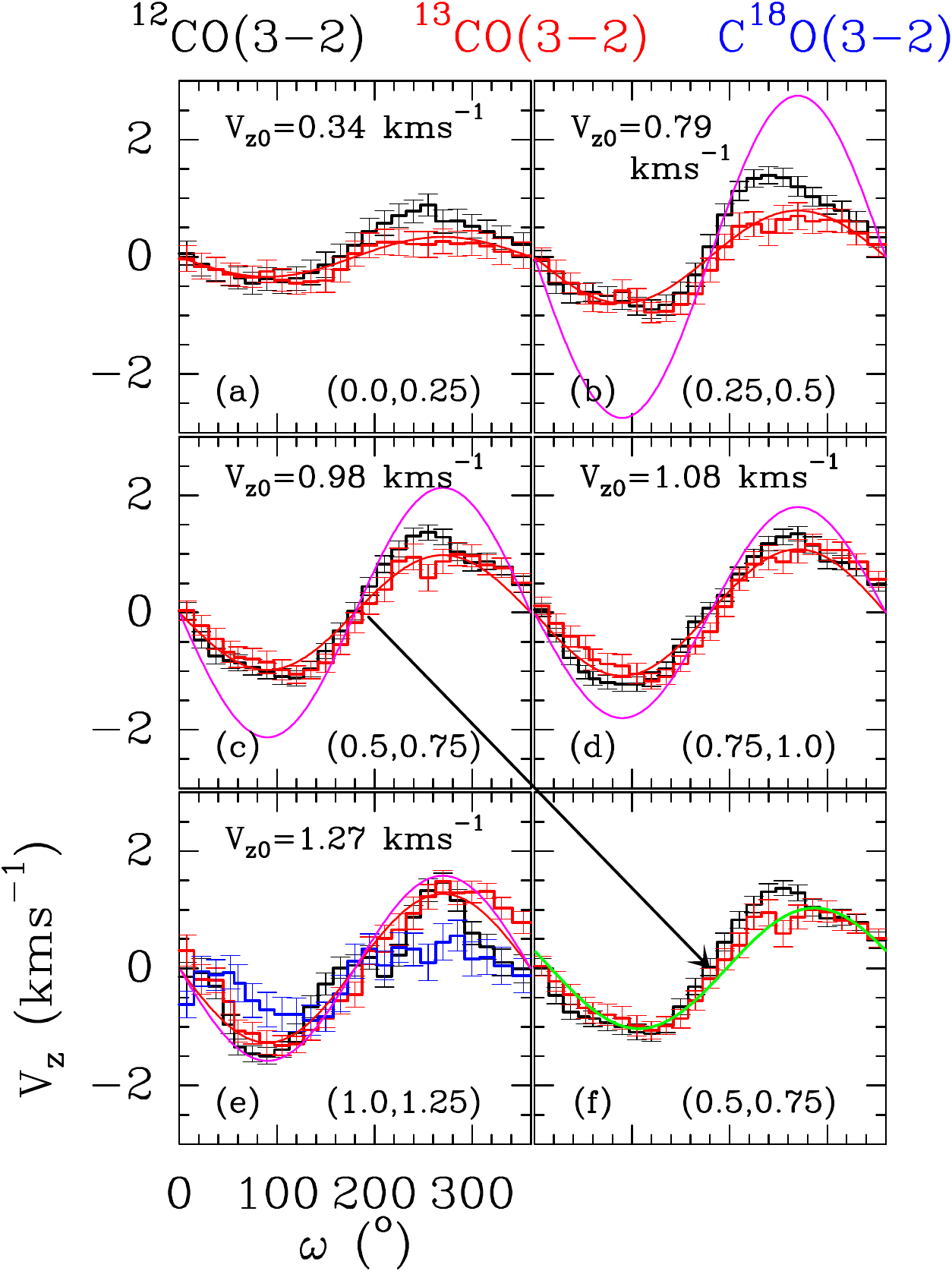}
  \caption{Dependence of $\langle V_z \rangle$ ($\kms$) on the 
  azimuth $\omega$ ($^\circ$) in the cavity. $^{12}$CO\,(3--2) is in 
  black, $^{13}$CO\,(3--2) in red and C$^{18}$O\,(3--2) in blue. The red 
  curve is a fit of a sine function to the $^{13}$CO\,(3--2) data (see 
  text). We use a blank space when no good data are available (no 
  emission) in the ring. The magenta curves show the expected Keplerian 
  velocity around a single star of 1.36\,M$_\odot$.
  The green curve in panel (f) shows the best-fit velocity curve 
  when infall motions are allowed, superimposed
  on the $^{13}$CO velocity.  
  } \label{fig:vzcav}
  
\end{figure}

To study the gas dynamics inside the cavity, we plot in 
Fig.\ref{fig:vzcav} the azimuthal dependence of $\langle V_z/\sin(i) 
\rangle$ in five rings of width $0.25''$ each of $^{12}$CO(3--2), 
$^{13}$CO(3--2), 
and \mbox{C$^{18}$O(3--2)} 
  emissions in 
the region $0<r<1.25''$. Azimuth and radius are defined in the disk 
plane, i.e. deprojected from the disk inclination. In each ring, we 
fit the azimuthal dependency of $\langle V_z/\sin(i) \rangle$ of the
$\tco$ with a sine function  \mbox{$V_z/\sin(i)=V_{z0}\sin\omega$}. 
This sine function is presented as the smooth red curve and the 
amplitude $V_{z0}$ is provided at the top of each panel. The good fit for 
the $\tco$(3--2) indicates that the gas inside the cavity is dominated by 
rotation. The amplitude is smaller than that of the Keplerian 
velocity 
$v(r) = \sqrt{G M_{tot}/r}$, however, where $M_{tot}$ is the total mass of the stars.
However, this is most likely a result of the finite 
resolution of the observations combined with the very 
inhomogeneous brightness distribution. The dynamics of the three lines 
agree for $1''<r<1.25''$, but differ in the region 
with $r<1''$. In particular, the $\dco$(3--2) departs from the $\tco$(3--2) 
in the region $0.25''<r<1''$ (boxes (b,c,d) in Fig.\ref{fig:vzcav}) 
because of the bright localized emission regions seen in CO. 

However, a better fit to the observed velocities is obtained by taking into account 
the contribution of a radial (from the disk center) velocity 
$V_z/\sin(i)=V_{fall}\cos\omega + V_{rot}\sin\omega$.
The results are 
presented in Table \ref{tab:vfall}: $V_{fall} > 0 $ corresponds 
to infall motions. As an example, the improvement in fit quality is
shown for the range $0.5''<r<0.75''$ in
panel (f) of Fig.\ref{fig:vzcav}. Table \ref{tab:vfall} thus indicates that the gas in the cavity 
moves inwards to the center at velocities about $0.3\,\kms$, 
which is about $10-15 \%$ of the Keplerian velocity. 
Because infall and rotation motions have different radial and azimuthal
dependencies, the finite beam size has a different effect on the 
infall velocity than on the apparent rotation velocity,
especially in the presence of brightness inhomogeneities.
  
\begin{table}  
\caption{Infall and rotation velocities of the gas inside the cavity.}
\begin{center}
\setlength{\tabcolsep}{4.5pt} 
\renewcommand{\arraystretch}{1.} 
\begin{tabular}{|c|c|c|c|c|c|}
\hline
 \multirow{2}{*}{Ring}   & $V_{Kep}$ & $V_{rot}$ & $V_{fall}$ & \multirow{2}{*}{$\frac{V_{fall}}{V_{rot}}$} & \multirow{2}{*}{$\frac{V_{fall}}{V_{Kep}}$} \\
           & ($\kms$) & ($\kms$) & ($\kms$) & &\\
\hline
(a) & - &0.34 &0.04 & 12\% & - \\
(b) & - &0.79 & 0.21 & 27\%  &- \\
(c) & 3.63&0.98 & 0.30 & 31\% &8\%\\
(d) & 3.07&1.08 & 0.38 & 28\% &12\%\\ 
(e) & 2.71&1.27 & 0.48 & 38\%  &18\%\\      
\hline                          
\end{tabular}
\end{center}
\label{tab:vfall} 
\vspace{-0.5cm}\tablefoot{The rings (a)...(e) are defined in Fig.\ref{fig:vzcav}}
\end{table}

\begin{figure} 
    \centering
     \includegraphics[height=5.0cm]{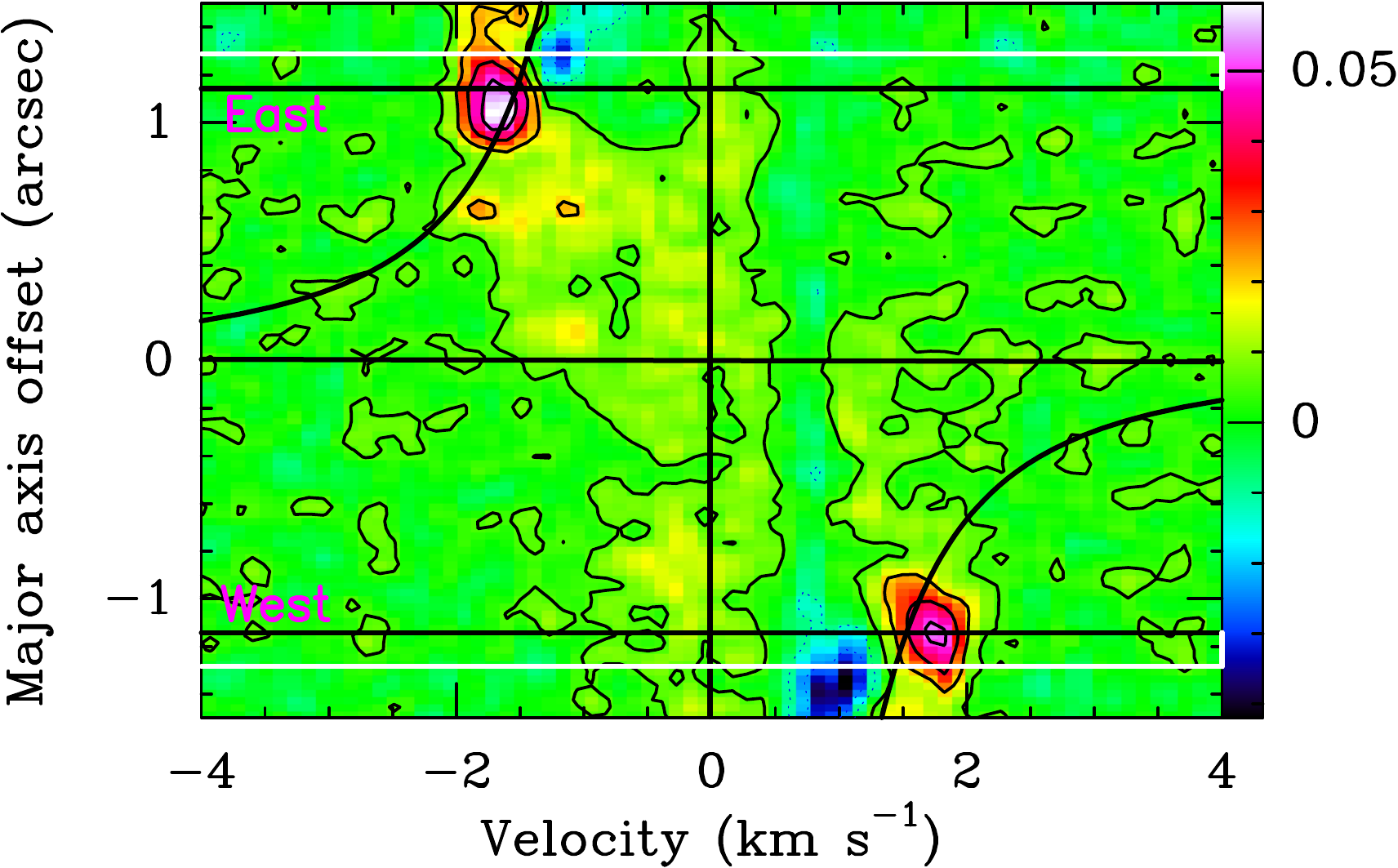}\\
     \vspace{.50cm}
     \includegraphics[height=5.0cm]{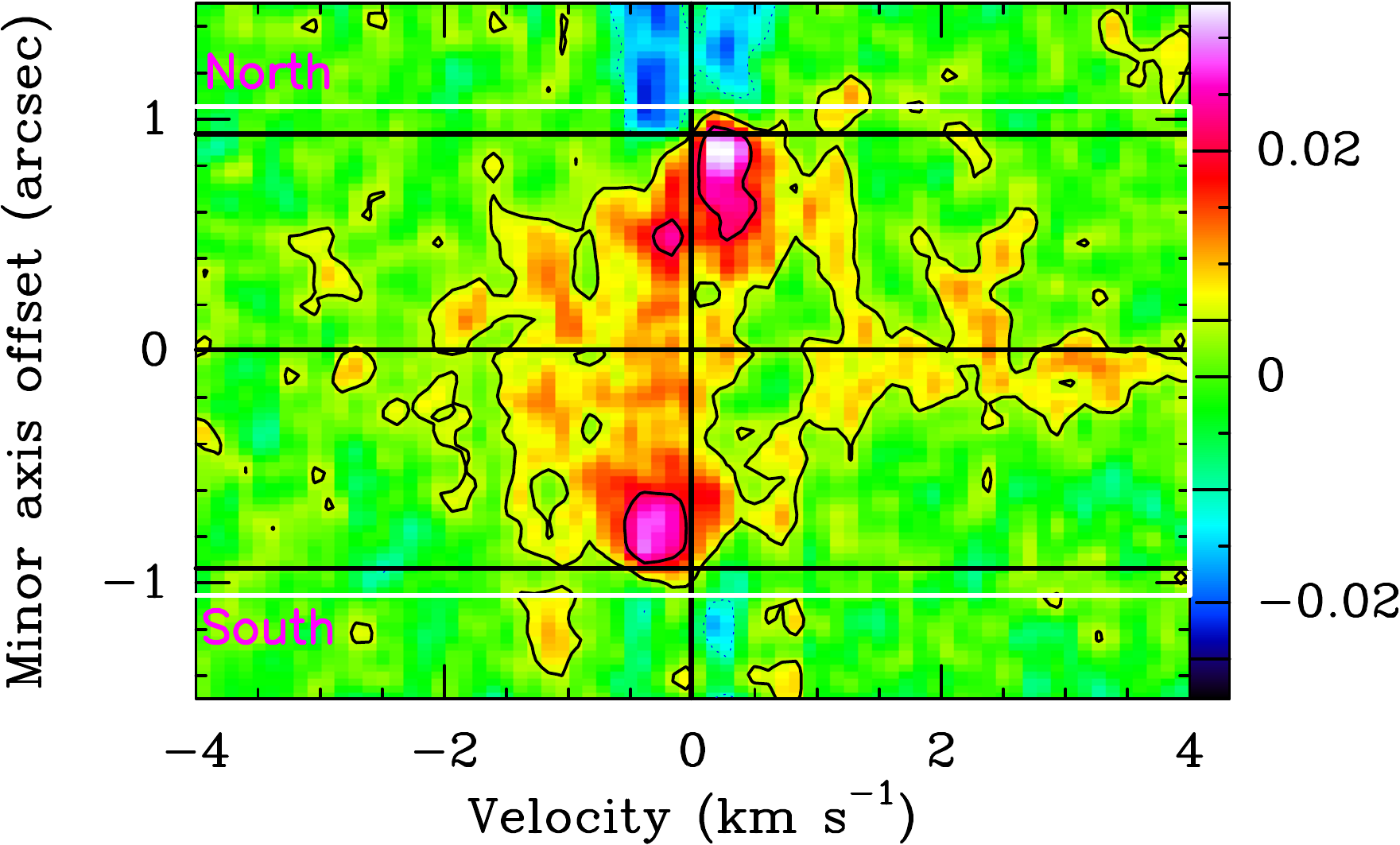}
  \caption{Position-velocity diagrams of the $\tco$(3--2) emission in 
  the cavity along the major (\textit{upper panel}) and minor axis (\textit{lower 
  panel}). The black curves show the expected Keplerian velocity around a 
  single star of  $1.36\Msun$. Contour levels are spaced by 10 mJy/beam;
  the zero contour is omitted. The white lines indicate the position
  of the inner edge of the dust ring (180 au) and the black lines show the 
  position of the inner radius of the gas disk (169 au). The data were
  rotated by $7^\circ$ to align with the disk axis, so that cardinal
  directions are approximate.}
  \label{fig:pv}
\end{figure}

A direct illustration of the infall motions is given in 
Fig.\ref{fig:pv}, which shows position-velocity (PV) diagrams of the 
$\tco$(3--2) emission in the cavity along the major and minor axis of 
the disk. The PV diagram along the major axis shows the Keplerian 
rotation down to the inner edge of the $\tco$(3--2) emission, at 
$\sim1.1''$ ($160$\,au). The PV diagram along the minor axis 
shows an asymmetry between the north and the south consistent with the 
derived infall motion of $\sim0.3-0.4\kms$ at the same 
(deprojected) radius (the PV diagrams are presented in the sky 
plane). 


\subsection{Gas properties}
\label{sec:nLTE}

\begin{figure}[!htp] 
    \centering
     \includegraphics[width=7.0cm]{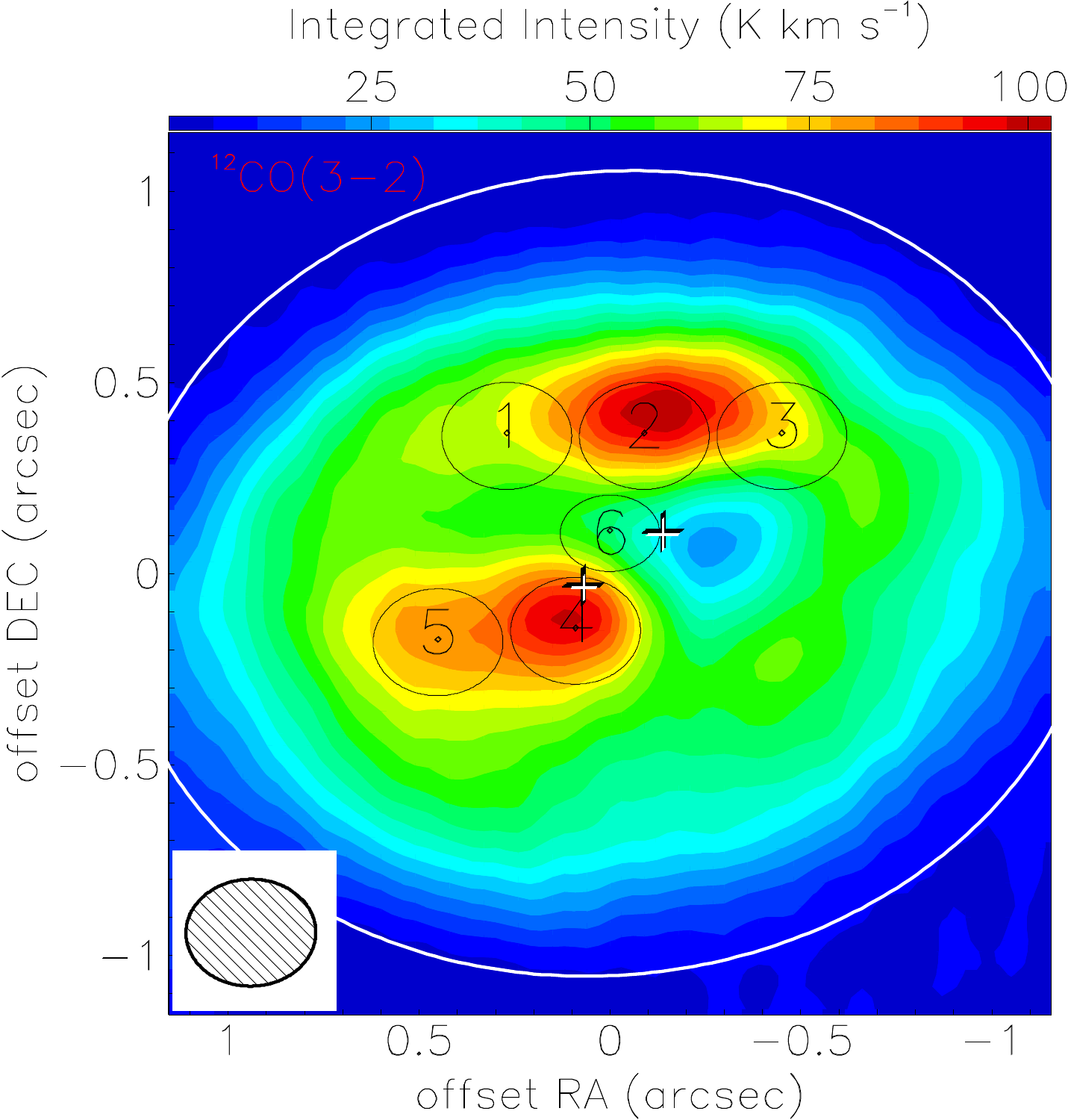}\\
     \vspace{.50cm}
     \includegraphics[width=7.0cm]{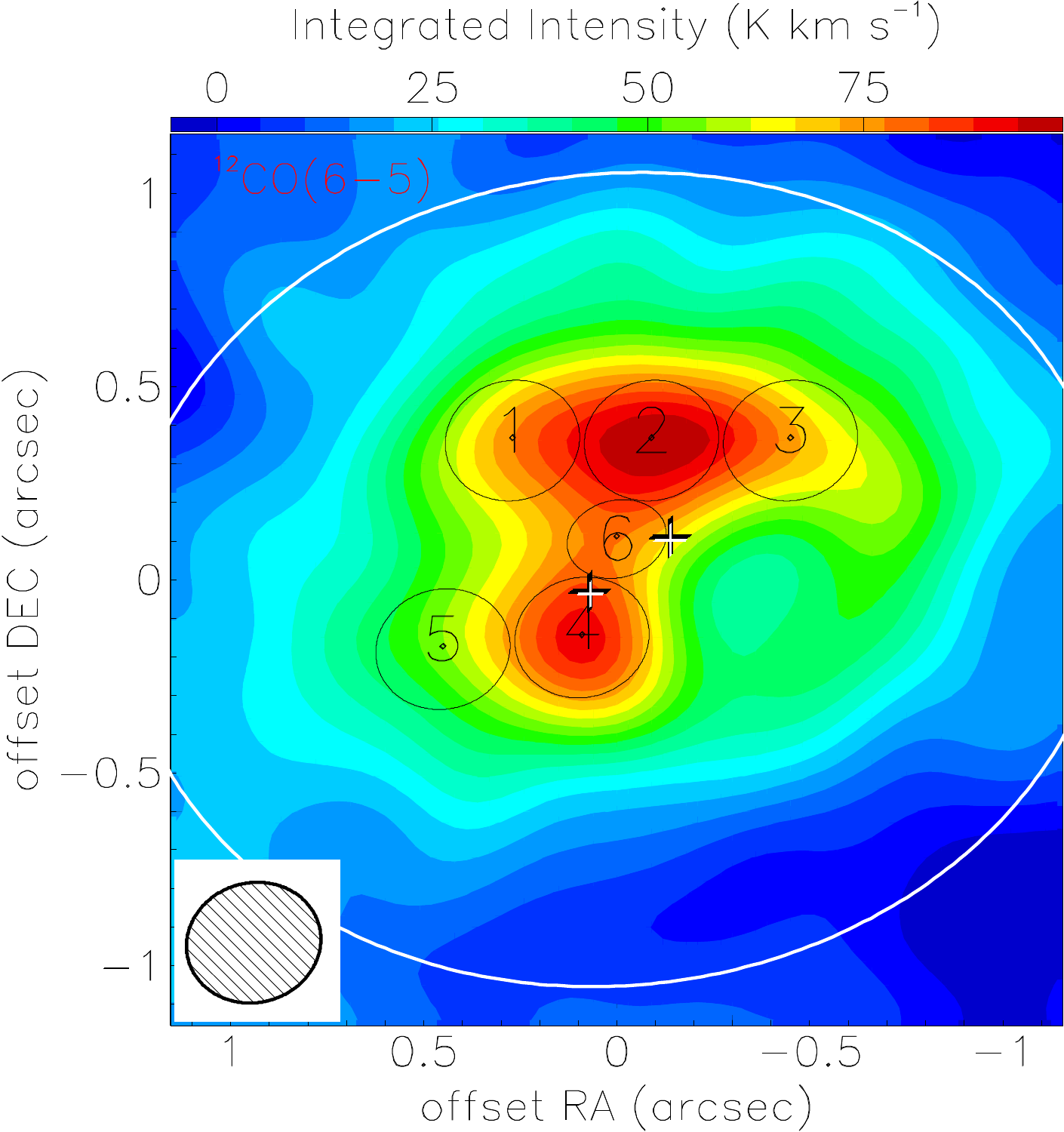}
\caption{Integrated intensity map of $\dco$(3--2) {\it{(upper)}} 
\citep[this work and][]{Tang+Dutrey+Guilloteau+etal_2016} and 
\mbox{$\dco$(6--5)} {\it{(lower)}} 
\citep[from][]{Dutrey+DiFolco+Guilloteau_2014} and positions and 
sizes of the blobs. The crosses mark the positions of Aa and Ab, and the ellipse
is the inner edge of the dust ring (180 au).}
   \label{fig:blobs}
\end{figure}

Using the (residual) $\dco$(3--2) 
data and the $^{12}$CO(6--5) data from 
\citet{Dutrey+DiFolco+Guilloteau_2014}, for which the emission
outside of the cavity is negligible, 
smoothed to a similar angular resolution ($0.35''\times0.30''$),
we identify 2 dominant features.
We decomposed these strong emission regions into five 
independent beams  (``blobs'') that we analyzed separately. To this we added 
a sixth  blob that connected blobs 2 and 4, a region that is 
particularly bright in CO(6--5) (see Figure \ref{fig:blobs}). Integrated line flux and line 
width were derived for each blob by fitting a Gaussian to the line 
profile. Velocities and line width derived from the CO\,(3--2) were
used to determine the $^{13}$CO and C$^{18}$O line intensities.

\begin{table*}
\caption{Properties of the brighter blobs}
\label{tab:blobs} 
\begin{center}
\setlength{\tabcolsep}{4.5pt} 
\renewcommand{\arraystretch}{1.5} 
\begin{tabular}{ccccccccccccc}
\hline
Blob  &  Position & Radius & $dv$ & H$_2$ density & N & T$_{kin}$ & Mass (nLTE) & Mass ($\dco$ Flux) & Mass ($\tco$ Flux)\\
      & ($'',''$) & ($''$) & ($\kms$) &(cm$^{-3}$) & (cm$^{-2}$) & (K) & ($\Msun$)& ($\Msun$)& ($\Msun$)\\
(1)  & ( 2) & (3) & (4) & (5) & (6) & (7) & (8) & (9) & (10) \\      
  \hline\hline
  1 & (0.27, 0.36)   &0.45 &2.5 &$>5.0\,10^4$   & $(2.1^{+0.6}_{-0.7})\, 10^{17}$ & $40\pm5$  &$(2.1^{+0.6}_{-0.7})\,10^{-6}$ & $(3.1\pm0.1)\,10^{-7}$ &$(1.9\pm0.1)\,10^{-6}$\\  
  2 & ($-$0.09, 0.36)&0.37 &2.7 &$>1.0\,10^4$ & $(1.4^{+0.7}_{-0.5})\, 10^{17}$ & $50\pm5$  &$(1.3^{+0.7}_{-0.5})\,10^{-6}$ & $(4.5\pm0.1)\,10^{-7}$ & $(2.3\pm0.1)\,10^{-6}$\\ 
  3 & ($-$0.45, 0.36)&0.58 & 2.1&$>5.0\,10^4$ & $(2.6^{+1.0}_{-1.00})\, 10^{17}$ & $40\pm5$  &$(2.5^{+1.0}_{-1.0})\,10^{-6}$ & $(3.2\pm0.2)\,10^{-7}$ & $(1.1\pm0.1)\,10^{-6}$\\ 
  4 & (0.09, $-$0.15)&0.17 & 6.2&$>1.0\,10^5$ & $(6.3^{+2.1}_{-1.4})\, 10^{16}$ & $80\pm10$ & $(6.3^{+2.1}_{-1.4})\,10^{-7}$& $(6.9\pm0.1)\,10^{-7}$ & $(2.5\pm0.2)\,10^{-6}$\\ 
  5 & (0.45, $-$0.18)&0.48 &2.5& $>1.0\,10^4$ & $(2.2^{+1.0}_{-0.8})\, 10^{17}$ & $40\pm5$ & $(2.2^{+1.0}_{-0.8})\,10^{-6}$& $(3.9\pm0.1)\,10^{-7}$ & $(1.7\pm0.1)\,10^{-6}$\\ 
  6 & (0, 0.1)       &0.12 &6.1& $>1.0\,10^4$      & $(4.0^{+1.4}_{-1.4})\, 10^{16}$ & $80\pm10$ & $(4.0^{+1.4}_{-1.4})\,10^{-7}$&$(3.4\pm0.1)\,10^{-7}$ & $(2.8\pm0.1)\,10^{-6}$\\
\hline                                   
\end{tabular}
\end{center}
\vspace{-0.5cm}
\tablefoot{ (1) Blob, (2) Offset from ring center, (3) Distance from 
center, (4) line-width  ($\kms$), (5) H$_2$ density, (6) CO column
density, (7) kinetic temperature, (8) H$_2$ mass derived from the CO column  
density (non-LTE analysis), (9) H$_2$ mass derived from the $\dco$ flux and (10) 
H$_2$ mass derived from the $\tco$ flux. } 
\end{table*}

To determine the physical conditions, we used a non-LTE escape 
probability radiative transfer code implemented in DiskFit. It uses the
escape probability formulation of \citet{Elitzur_1992}, 
\mbox{$\beta=[1-\exp(-\tau)]/\tau$}, a single collision partner, H$_2$, and 
Gaussian line profiles. 
Non-LTE best-fit solutions were found by sampling the $\chi^2$ 
surface, which is defined as the quadratic sum of the difference between the 
measured brightness temperatures and the computed values of the CO\,(6--5), CO\,(3--2),
$^{13}$CO\,(3--2), and C$^{18}$O\,(3--2) transitions, for ranges 
of H$_2$ density of $10^2-10^{10}$\,cm$^{-3}$, $\dco$ column density of 
$10^{13}-10^{19}$\,cm$^{-2}$, and kinetic temperature of $3-100$\,K 
using 50 steps of each parameter. We assumed the standard isotopic 
ratios $^{12}$C/$^{13}$C$=70$ \citep{Milam+Savage+Brewster_2005} and 
$^{16}$O/$^{18}$O$=550$ \citep{Wilson_1999} for the relative abundances 
of the isotopologs. $\dco$ constrains the temperature, and $\tco$ the 
column densities. Owing to its faintness, the C$^{18}$O(3--2) data 
bring little information. Because the critical densities 
of the observed transitions are low, we only obtain a lower limit to the 
density. The blob parameters are presented in Table \ref{tab:blobs}.

We typically find high CO column densities of about  a few 
$\sim10^{17}$\,cm$^{-2}$ and temperatures in the range 40--80\,K, with a 
lower limit on the density  of about 10$^{5}$\,cm$^{-3}$.
For blob 6, the faintest region analyzed with this method, the
problem is marginally degenerate, with two separate solutions: i)
a high column density ($\sim10^{17}$\,cm$^{-2}$) and low temperature 
($\sim20$\,K), and (ii) a low column density 
($\sim10^{15}$\,cm$^{-2}$) and high temperature ($>80$\,K).
This region lies between Aa and Ab, therefore the second solution (which is
also that of lowest $\chi^2$) is more probable. 

\subsection{Gas masses}
\label{sub:mass}

The lower limit on the density obtained from the non-LTE
analysis was insufficient to provide any useful constraint on the
blob masses, we therefore used another method.  We estimated the blob
mass from the derived molecular column density and blob size, 
assuming a molecular abundance relative to H$_2$, as described below.

In the same way, 
we also derived the total amount of gas in 
the cavity from the integrated flux of the optically thin lines of the 
$^{13}$CO(3--2) and C$^{18}$O(3--2). For this purpose, we integrated
the emission out to a radius of 160\,au. 

In the optically 
thin approximation, the integrated flux and the column density of the 
upper level of a given transition are related by
\begin{equation}
W=\frac{g_u}{\gamma_u}N_{u} \\\\
\end{equation}
where $W=\int{T_b\,dv}$ is the integrated brightness 
inside the cavity ($R<160$\,au), $g_u=2J+1$ is the statistics weight 
and $N_u$ is the column density of the upper level, 
$\gamma_u=\frac{hc^3A_{ul}}{8\pi\,k_B \nu^2}$ (the Einstein coefficient 
$A_{ul}$ is taken from Lamda database\footnote{https://home.strw.leidenuniv.nl/~moldata/}).  
Guided by the results of the non-LTE analysis, 
we assumed that the gas temperature $T$ is 40\,K everywhere inside the cavity 
and calculated the total column density $N_{\textnormal{total}}$ of a 
given molecule as 
\begin{equation}
N_{\textnormal{total}}=\frac{N_u}{Z}\exp\left(\frac{-E_u}{k_B\,T}\right)
\end{equation} 
where, $Z$ is the partition function and $E_u$ is the energy of the upper state.
The $^{12}$CO(3--2) emission, which is  
partially optically thick, yields a lower limit.

The CO abundance was 
taken from the abundances measured in TMC-1 by 
\citet{Ohishi+Irvine+Kaifu_1992}, and we assumed a standard isotopic
ratio for the isotopologs ($\tco$ and $\cdo$).
The results are listed in Cols 9-10 of Table \ref{tab:blobs} for the
blobs, and 
Table \ref{tab:mass-cavity} summarizes the results for the
cavity. 

\begin{table}[!htp]
\caption{Mass of gas inside the cavity}
\label{tab:mass-cavity} 
\begin{center}
\setlength{\tabcolsep}{4.5pt} 
\renewcommand{\arraystretch}{1.} 
\begin{tabular}{cccc}
\hline
Location  &  Integrated Flux & H$_2$ mass  & Abundance  \\
      & (Jy\,$\kms$) & ($\Msun$) & (w.r.t H$_2$)\\
\hline
  Cavity ($\dco$) & $11.4\pm0.8$& $6.1\pm0.4\times10^{-6}$ & $8.0\times10^{-5}$\\
    Cavity ($\tco$) & $3.8\pm0.1$& $1.6\pm0.1\times10^{-4}$ & $\dagger$\\
      Cavity ($\cdo$) &$0.5\pm0.2$&$1.6\pm0.8\times10^{-4}$ &$\ddagger$\\
\hline                               
\end{tabular}
\end{center}
\vspace{-0.5cm}
\tablefoot{$^\dagger$ X[$\tco$]=X[$\dco$]/70 and $^\ddagger$ X[$\cdo$]=X[$\dco$]/550 (see text).} 
\end{table}

The H$_2$ mass derived from $\tco$(3--2) and C$^{18}$O(3--2) is similar 
which confirms that these lines are optically thin while the $\dco$ 
emission is optically thick. 
\section{Discussion}
\label{sec:disc}

\begin{figure}[h] 
  \centering
  \includegraphics[width=\columnwidth]{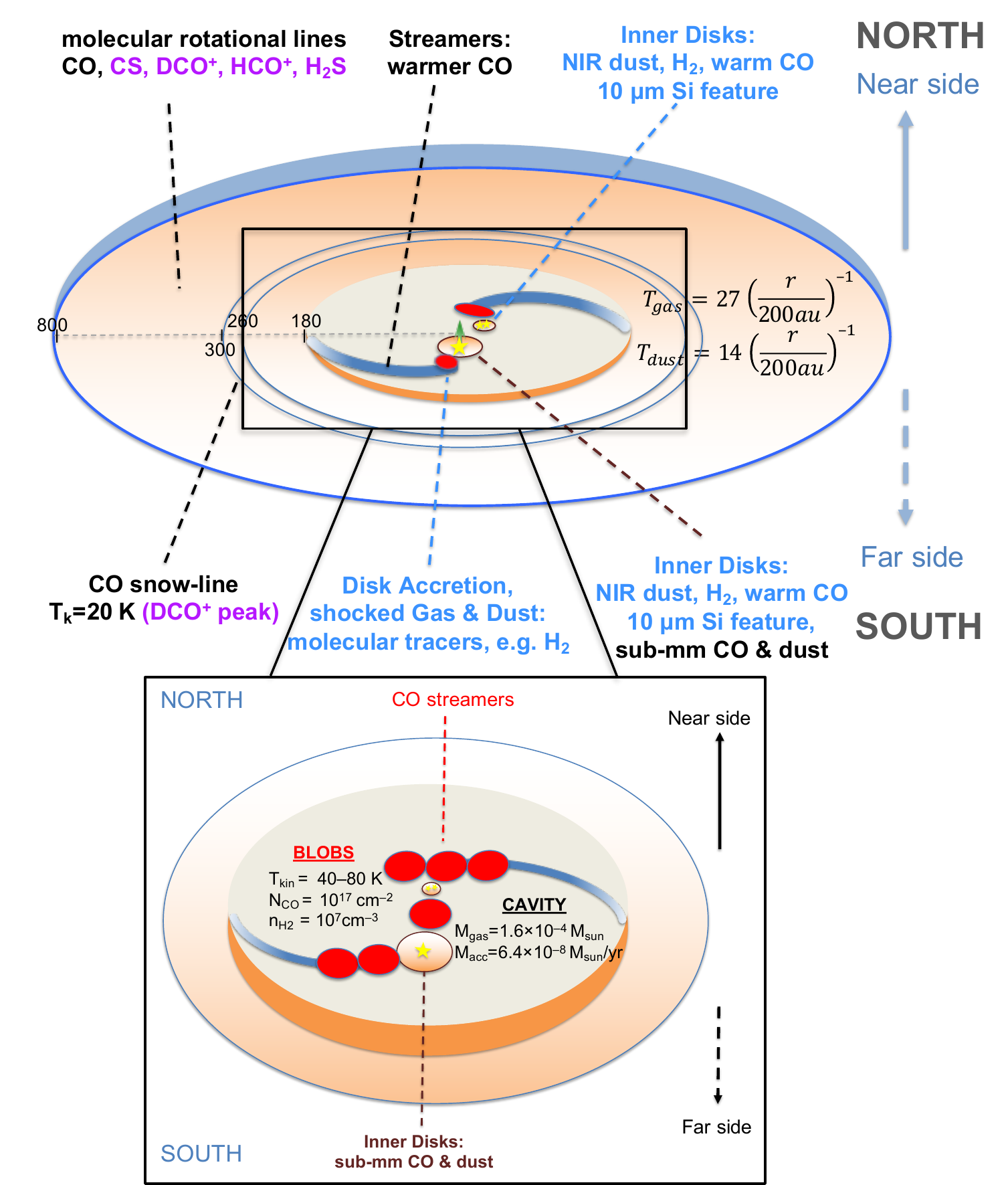}
  \caption{Schematic view of the GG Tau system. Colored text indicates
  results obtained from other publications. In particular, magenta shows 
  results from 
  \citet{Phuong+Chapillon+etal_2018}; see  
  \citet{Dutrey+DiFolco+Beck_2016} for a review of other references.
  Black text shows results from this work.}
  \label{fig:scheme}
\end{figure}

Figure \ref{fig:scheme}  is a schematic layout summarizing the properties
of the GG Tau A system. Numbers quoted in this schematic view are discussed
in the following section.

\subsection{Temperature distribution in the outer disk}

Our analysis confirms  that most of the outer disk of GG Tau A is very 
cold (see Fig.\ref{fig:cotemp}). The gas temperature derived here agrees with the 
value found by \citet{Guilloteau+Dutrey+Simon_1999} from $\tco$ 
alone, 20\,K at 300\,au. The agreement between values derived 
independently from $\dco$ and $\tco$ supports our assumption of a 
limited vertical temperature gradient in the CO layer, as already 
mentioned by \citet{Tang+Dutrey+Guilloteau+etal_2016}. A power-law fit 
to this temperature profile gives a radial dependency of $r^{-1} - 
r^{-1.3}$, confirming the previous exponent value  of $-1$ derived by 
\citet{Guilloteau+Dutrey+Simon_1999}.
\citet{Dutrey+DiFolco+Guilloteau_2014} also found a similar exponent 
for the dust temperature from the analysis of dust images between 3 and 
0.45\,mm using a simple power law. Because this study is based on 
multi-wavelength continuum resolved observations from 3 up to 0.45\,mm, 
the derived dust temperature is characteristic of the large grains 
that have likely settled down around the midplane. It is then 
reasonable to consider that this temperature traces the midplane dust 
temperature. This steep radial slope of the temperatures is most likely 
due to the stellar light that is blocked by the inner dense ring, and 
the rest of the disk then remains in its shadow. 

\subsection{Gas distribution and smoothness of the outer ring}

\paragraph{Global properties}
Our canonical (smooth) model (Sec.\ref{sec:sub:coiso}) shows 
that the ratio of the $\tco$ and $\cdo$ column densities beyond 
\mbox{$r > 300$\,au}  is about 17 (see Table 
\ref{tab:co-final}) above the standard isotopic ratio of 7, suggesting 
selective photodissociation 
\citep[e.g.,][]{vanDishoeck+Black_1988}, but also confirming 
that in the outer disk the emissions are optically thin. 

In contrast, inside the densest part of the ring \mbox{($200-260$\,au)}, 
the measured ratio is about $3-5$. Chemical effects
such as selective photodissociation and fractionation which occurs 
through isotope exchange between CO and C$^+$ \citep{Watson+etal_1976}
and enhances $^{13}$CO at temperatures of about \mbox{20-30}\,K, would both tend
to enhance this ratio above the double isotopic 
($^{16}$O/$^{18}$O)/($^{12}$C/$^{13}$C) ratio. Thus the simplest 
explanation for a low value is partially optically thick $\tco$(3--2) 
emission. However, our model should account for the opacity if the disk 
were as smooth as assumed. We therefore conclude that the GG Tau disk and 
ring deviate significantly from the smooth, nonstratified, 
azimuthally symmetric structure we adopted.

\paragraph{Smoothness versus unresolved structures}
The radial profile (see the upper panels in  Fig.\ref{fig:profile}) of the 
observed molecular lines, $\dco$(3--2), $\tco$(3--2), and C$^{18}$O(3--2) 
does not appear smooth. After subtracting the best 
(smooth) outer disk model, Figure \ref{fig:residual} reveals some additional  
emission that is located in rings at specific azimuths. This is particularly 
clear for the optically thinner transitions of the C$^{18}$O(3--2), 
suggesting radial density variations in the molecular layer 
(at about one scale height). In contrast to the gas, the dust emission is 
hardly visible in the outer disk (radius $>260$\,au), but mostly 
concentrated in the ring (radius $200-260$\,au). 

The azimuthal dependence of the integrated brightness of the 
$^{12}$C$^{16}$O emission (lower left panel in Fig.\ref{fig:profile}) shows 
strong excesses at specific azimuths. The
excess seen in the south-east quadrant is consistent 
with the hot-spot location quoted by 
\citet{Dutrey+DiFolco+Guilloteau_2014}. This  hot 
spot remains visible, but less clearly so in $\tco$ 
and $\cdo$. This indicates that it is 
mostly a temperature enhancement, and not an overdensity region.

In the residual maps (Fig.\ref{fig:residual}), other azimuthal 
variations are also visible. Our smooth model removes any azimuthally 
symmetric emission so that 
apparent effects resulting from velocity coherence length and convolution 
with elongated beam-shape are properly eliminated. 
The observed residuals thus reveal intrinsic structures.

All this evidence indicates the existence of radial and 
azimuthal substructures that remain unresolved (at least radially)
at our 30\,au linear resolution.

\subsection{Properties of the gas inside the cavity} 
\label{sec:sub:cavity}

\paragraph{Kinematics}
Figure \ref{fig:vrot} shows that the rotation appears sub-Keplerian at 
radii smaller than about $0.8''$. This could be the signature of the
tidal forces generated by the Aa/Ab binary. 
Unfortunately, this is largely an effect of 
the intensity drop in the cavity, combined with the finite angular 
resolution. Because the signal intensity increases with radius in the 
cavity, the intensity-weighted mean velocity is biased towards the 
values obtained at the largest radii, i.e. the gas apparently rotates 
at lower velocities.  A proper modeling of the angular resolution 
effect, accounting for the observed brightness distribution, would be 
required to remove this artifact and determine whether the gas rotate 
at the expected Keplerian speed.

On the other hand, we find clear evidence for infall motions in the 
cavity (see Sec.\ref{sec:dynamics}), at velocities about $10-15\%$ of 
the Keplerian speed, proving that material is accreting onto the inner 
disks orbiting the central stars. This is consistent with the infall 
value found for L 1551 NE, a younger binary system 
\citep{Takakuwa+etal_2017}. However, our sensitivity is insufficient for a
detailed comparison with hydrodynamics models.

In summary, we find that the gas starts to exhibit non-Keplerian 
motions (at least infall motions, and perhaps slower than 
Keplerian rotation) at $r \approx 160$\,au, somewhat smaller than the 
inner edge of the dust ring (193 au).  This difference in
radii is expected when dust trapping in the high-pressure bump
that occurs in the dense ring is considered \citep[e.g.][]{Cazzoletti+etal_2017}.
The 160\,au radius remains much larger 
than the radius at which tidal disturbances are expected in a binary 
system, however. Theoretical studies showed that such disturbances are expected at radius of $2.5-3$ times 
the major axis of the binary orbit \citep{Artymowicz+Lubow_1996}. The binary separation  
of Aa/Ab, 35 au, means that this theoretical value should be of the order of 100 au. 
Therefore, we would expect that deviations 
from Keplerian motions only appear inside about 100\,au, unless the orbit is very 
eccentric. High eccentricity appears unlikely given the measured 
orbital parameters; see \citet{Beust+Dutrey_2005} who also
reported that underestimated astrometric error bars might play
an important role. Following \citet{Beust+Dutrey_2005}, \citet{Koehler_2011} and 
\citet{Nelson+Marzari_2016} showed that this apparent contradiction 
could be solved under the assumption that the orbital plane of the stars is 
very different from the (common) plane of the ring and outer disks. A 
similar result was found by \citet{Aly+Lodato+Cazzoletti_2019}, who 
indicated that an inclination difference of 30$^\circ$ could remain 
stable over the (circumbinary) disk lifetime. However, 
\citet{Brauer+etal_2019} found by studying the projected shadows
of the disks in the near IR,
that the circumstellar disk around Aa and one 
of the disks around Ab1 or Ab2 must also be coplanar with the 
circumbinary ring and disk. This makes the misaligned orbit proposition 
unlikely, because the alignment of the circumstellar disks is more
controlled by the gravitational interactions with the stars than
with the (much less massive) outer disk. The cavity size puzzle 
thus remains.

\paragraph{Gas temperature}
Our non-LTE analysis, in agreement with the study from 
\citet{Dutrey+DiFolco+Guilloteau_2014}, shows that the gas inside the 
cavity is warm, with temperatures ranging from 30 to 80\,K. In the 
bright blobs, near the stars, we derived a kinetic temperature of about  
 $40-50$\,K at about $30-60$\,au from the central stars. It is 
important to mention that these temperatures are well above the CO 
freeze-out temperature. 

\paragraph{Amount of gas}

From our non-LTE analysis of the bright blobs, we found a few 
$10^{17}$ cm$^{-2}$ for the CO column density with the exception of 
blobs 4 and 6,  which have a lower column density of $\sim$(3-6)\,
10$^{16}$\,cm$^{-2}$. We also obtained a lower limit on the H$_2$ 
density  of about  $(1-10)\,10^4$ cm$^{-3}$ for all blobs.   
However, a more stringent constraint can be obtained
from the blob masses given in Table \ref{tab:blobs}, because 
the thickness of the blobs is on the order of the scale height 
$H(r)$, 5 to 10\,au at this distance to the stars.
This leads to densities about $10^7$\,cm$^{-3}$.

The cumulative mass of the blobs is  $\sim1.2 
\times10^{-5}\Msun$. We also estimated the total gas mass inside the 
cavity from the integrated flux of the optically thin CO isotopologs. 
We found a mass of $\sim1.6\times10^{-4}\Msun$, assuming standard CO 
abundance (see Table \ref{tab:mass-cavity}). The $\tco$ and $\cdo$ values perfectly agree 
suggesting that both the $\tco$ and $\cdo$ emissions are essentially 
optically thin in the cavity. 

Therefore, the total mass of the gas inside the cavity appears to be a factor 
10 higher than the cumulative blob mass. This only relies on 
the assumption of similar molecular abundances in these regions, which 
is reasonable given their similar temperatures. Thus a significant 
fraction of the gas in the cavity does not reside in the dense blobs but in 
diffuse features. 

Determining the absolute value of the gas mass inside the 
cavity is more challenging. On the one hand, our assumed value for the CO 
abundance, that observed in TMC-1, appears reasonable given the relative high 
temperature in the cavity. However, lower values might result from C 
and O still being locked on grains in the form of more complex or less 
volatile molecules (CO$_2$ and CH$_4$, see \citet{Reboussin+etal_2015}). 
A proper quantification of such a process would require a complete 
chemical study that would follow  the physical and chemical evolution of the gas 
and solid phases throughout the disk. 

Nevertheless, an absolute minimum value for the gas mass in 
the cavity can be obtained when we assume the CO abundance cannot exceed 
the carbon cosmic abundance expected in cold molecular clouds 
\citep[$3.4\times10^{-4}$, ][]{Hincelin+etal_2011}. In this case, we obtain
the minimum mass by correcting the previous value by the factor of $\sim 0.2$. This leads to 
\mbox{$\sim0.3\times10^{-4}\Msun$} for the total gas mass inside the cavity.

In any case, the mass of gas in the cavity is only a very small 
fraction of the total disk mass (0.15$\Msun$) that is estimated from 
the dust emission.   

\paragraph{Mass accretion rate}

The gas in the cavity is unstable and will accrete onto the GG Tau A 
disks on a timescale of a few (about four) times the orbital binary period, 
which is estimated to be about 600 years \citep[see][]{Beust+Dutrey_2005}. 
A similar timescale, about 2500 yr, is given independently by the 
ratio of cavity radius to the measured infall velocities. This 
gives an accretion rate of $\sim6.4\times10^{-8}\Msun\,\rm{yr}^{-1}$ when 
we assume the canonical mass value. 
The accretion rate on GG Tau Aa+Ab, measured in year 2000 using 
the H$_\alpha$ 
line, is about $\sim2\times10^{-8}\Msun\,\rm{yr}^{-1}$  \citep{Hartigan+Kenyon_2003}, 
a factor 3 lower than our estimate. 
This small difference (a factor 3) may be essentially due to the limited accuracy of the 
two methods in use. However, it may also be partly explained by variable accretion inside 
the cavity and onto the central star(s) associated to non steady-state 
dynamics. In a binary star, the accretion rate process is modulated by 
the eccentricity, and is more efficient at the pericenter, with a delay 
that  depends on the eccentricity 
\citep{Artymowicz+Lubow_1996, Gunther+Kley_2002}. The two values of the 
accretion rates may then reflect different aspects of a highly variable process 
depending how and when these rates are measured. The fair 
agreement between the two results shows that the GG Tau A disk can be 
sustained by accretion through the cavity on a long timescale.
\section{Summary}
\label{sec:sum}
We reported new observations of CO isotopologs with ALMA of 
the close environment of  GG Tau A. We studied the ring by performing an LTE 
analysis, and we performed a non-LTE analysis for the gas clumps observed 
inside the cavity. We also investigated the gas kinematics in the outer 
disk and inside the cavity.  
 
The ring and outer disks do not exhibit a smooth distribution, but 
likely consist of a series of unresolved substructures with some 
azimuthal variations, particularly in the dense inner ring. The bright 
hot spot seen in $^{12}$CO is marginally seen in $\tco$ and in $\cdo$,  
suggesting a temperature effect.

The gas temperature derived from the optically thick CO line has 
a sharp decrease ($r^{-1}$), as for the dust. 
The temperature of 20\,K (CO snow line) is reached at $\sim$ 300 au.    

The total amount of mass inside the cavity derived from $\tco$ is 
$1.6\times10^{-4}\Msun$, assuming standard CO abundance. 

The gas streamers inside the cavity can essentially be defined by six
blobs. A non-LTE analysis reveals that their conditions are 
similar, with CO column densities of about a few 
$\sim10^{17}$\,cm$^{-2}$, temperatures in the range $40-80$\,K, and
H$_2$ density in the dense parts of about 10$^{7}$\,cm$^{-3}$.

The kinematics of the whole structure (outer ring plus cavity) 
appears to be in Keplerian rotation around a 1.36 $\Msun$ system for radii
beyond $\sim$1.2$''$ or 180\,au. The kinematics of the gas streamers and 
blobs appears more complex than expected for such a binary system. 
In particular, the gas starts to exhibit non-Keplerian motions
for radii smaller than $\sim 160$\,au.

The gas inside the cavity shows infall motions of 
about 10\% of the Keplerian velocity, allowing the central stars to 
accrete material from the dense ring.

The average mass flow rate of the gas through the cavity is  
$\sim6\times10^{-8}\Msun\,\rm{yr}^{-1}$. This value is compatible with the stellar 
accretion rate measured using the H$_\alpha$ line, and it is sufficient to
replenish the circumstellar disks.


\begin{acknowledgements}
N. T. Phuong warmly thanks P. Darriulat  for his guidance and supports. A. Dutrey also thanks him for making the collaboration possible. 
A. Dutrey and S. Guilloteau acknowledge M. Simon who started to study this wonderful object with them in another millennium.
This work was supported by ``Programme National de Physique Stellaire'' (PNPS)
and ``Programme National de Physique Chimie du Milieu Interstellaire'' (PCMI)
from INSU/CNRS.
This research made use of the SIMBAD database,
operated at the CDS, Strasbourg, France.
This paper makes use of the following ALMA data:
   ADS/JAO.ALMA\#2011.1.00059.S, \#2012.1.00129.S and \#2015.1.00224.S ALMA is a partnership of ESO 
   (representing its member states), NSF (USA), and NINS (Japan), together with NRC
   (Canada),  NSC and ASIAA (Taiwan), and KASI (Republic of Korea) in
   cooperation with the Republic of Chile. The Joint ALMA Observatory is
   operated by ESO, AUI/NRAO, and NAOJ.
N. T. Phuong and P. N. Diep acknowledge financial support from World Laboratory,
Rencontres du Viet Nam, the Odon Vallet fellowships, Vietnam National Space
Center, and Graduate University of Science and Technology. N. T. Phuong thanks the financial support 
from French Embassy Excellence Scholarship Programme, P. Darriulat and from the Laboratoire d'Astrophysique de Bordeaux.  
This research is funded by Vietnam National Foundation for Science and Technology Development (NAFOSTED) under grant number 103.99-2018.325.
\end{acknowledgements}
%

\begin{appendix} 
\onecolumn 
\newpage
\section{Channel maps} 
\begin{figure*}[h]
\includegraphics[width=\textwidth]{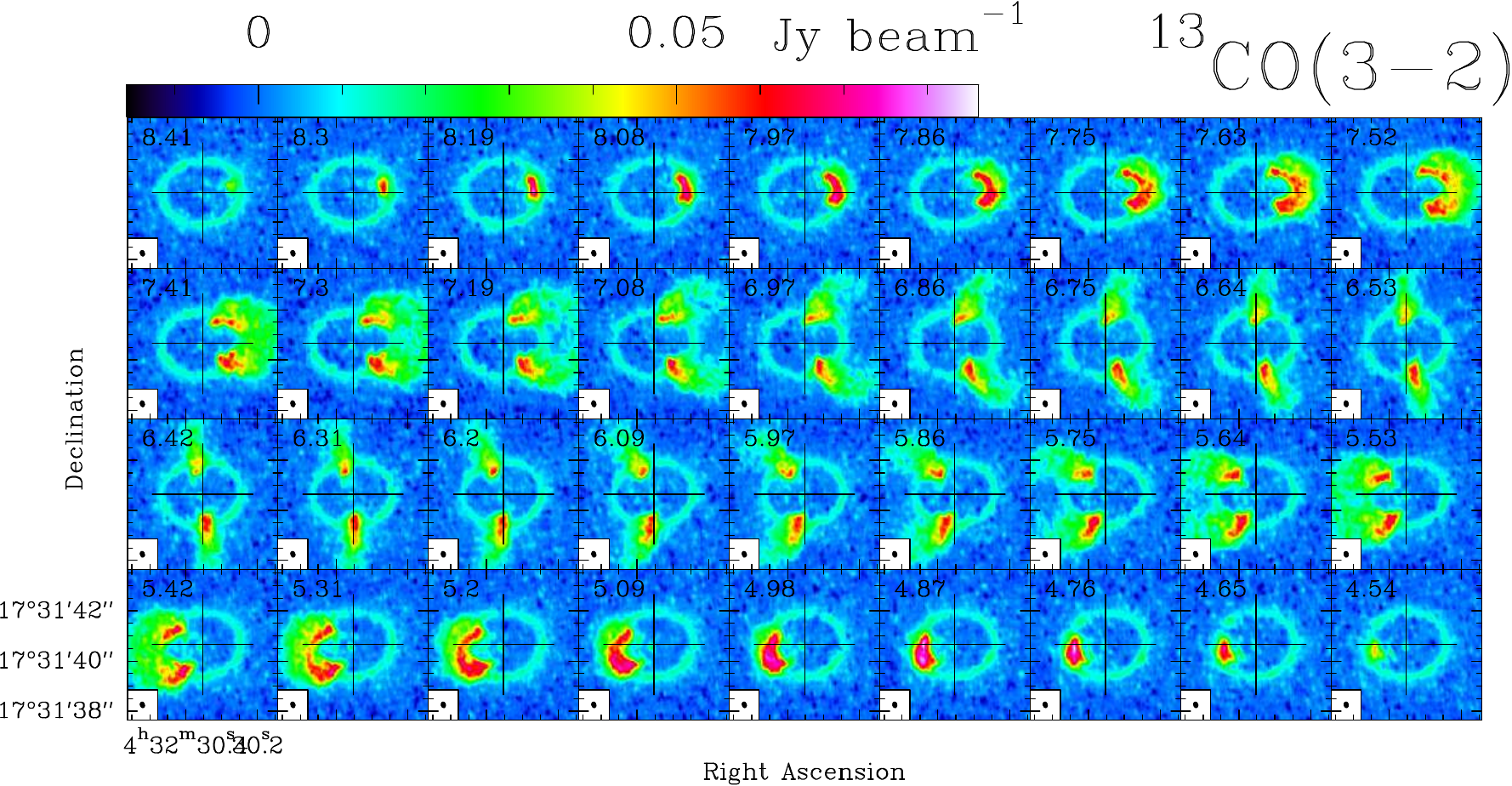}
 \caption{Channel maps of $\tco$(3--2), beam $0.22"\times0.16"$, 
 PA=$16^\circ$. The noise level is 2.4 mJy\,beam$^{-1}$. The color 
 scale is indicated in the upper panels. The cross is centered on 
 the center of the map.}
   \label{fig:13co32new}
\end{figure*}

\begin{figure*}[h]
\includegraphics[width=\textwidth]{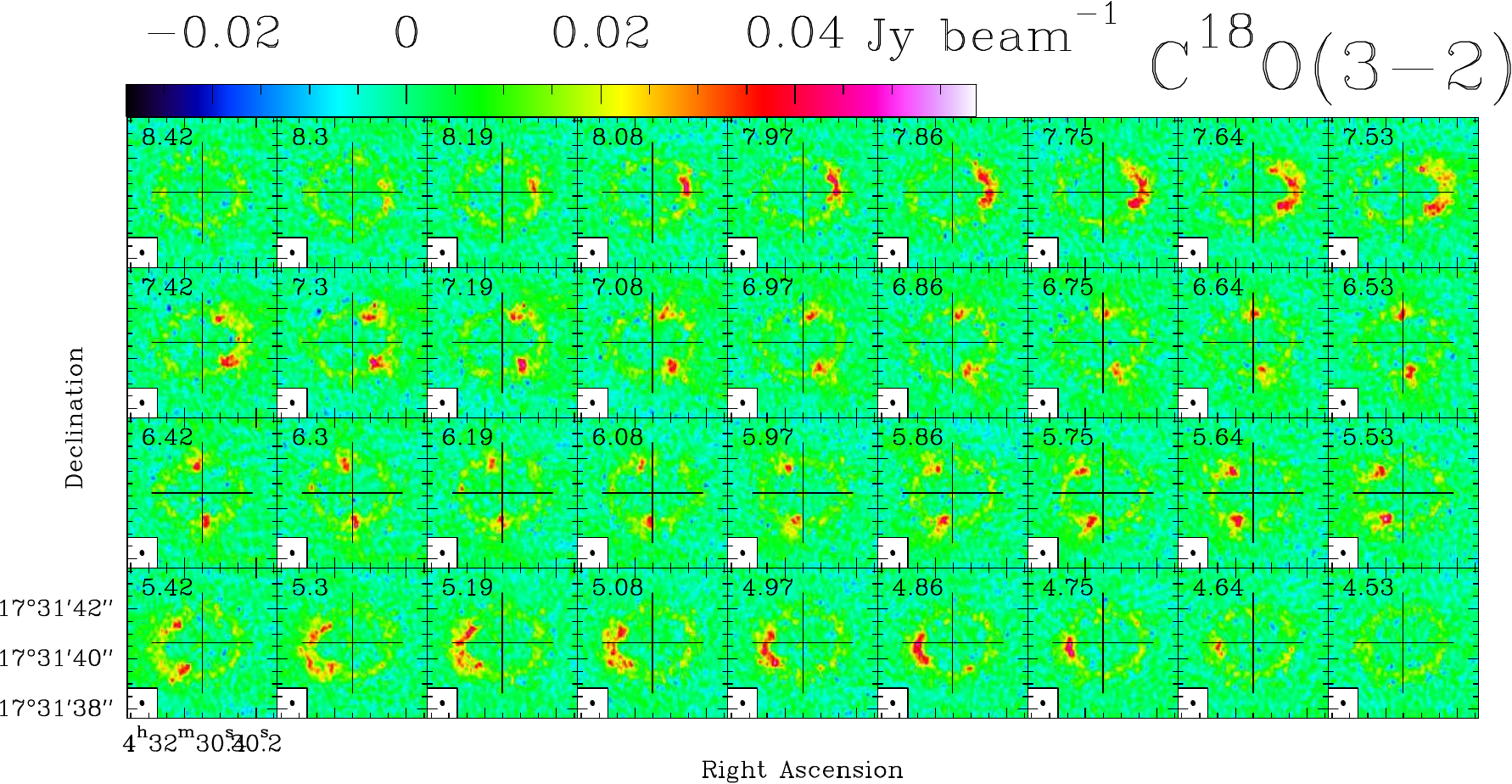}
 \caption{Channel maps of C$^{18}$O(3--2), beam $0.19"\times0.14"$, 
 PA=$19^\circ$. The noise level is 4.8 mJy\,beam$^{-1}$. The color 
 scale is indicated in the upper panels. The cross is centered on 
 the center of the map.} \label{fig:c18o32new}
\end{figure*}

We present in Figs.\ref{fig:13co32new}-\ref{fig:c18o32new} the Cleaned channel maps produced without subtracting
the continuum emission.
\label{app:chan}
\section{Disk model fitting}
\label{app:diskfit}

We used the tool DiskFit \citep{Pietu+Dutrey+Guilloteau+etal_2007} to 
derive the physical
parameters of rotating circumstellar disks.

\paragraph{Principles:}

DiskFit computes the spatial distribution of the emission 
from spectral lines (and dust) as a function of frequency (related
to the line rest frequency and Doppler velocity of the source) for a
given azimuthally symmetric disk model.  

In its basic form, as described in \citet{Pietu+Dutrey+Guilloteau+etal_2007}, the disk 
model assumes that the relevant physical quantities that control the 
line emission vary as power law as a function of radius, and, except for 
the density, do not depend on height above the disk plane. The exponent 
is taken as positive when the quantity decreases with radius:
\begin{equation}
 a(r) = a_0 (r/R_a)^{-e_a}.
\end{equation}

When dust emission is negligible, the disk for each molecular line is
thus described by the following parameters:
\begin{itemize}\itemsep 0pt
 \item $X_0, Y_0$, the star position, and $V_\mathrm{disk}$, the
 systemic velocity.
 \item PA, the position angle of the disk axis, and $i$ the
 inclination.
 \item $V_0$, the rotation velocity at a reference radius $R_v$, and
 $v$ the exponent of the velocity law. With our convention, $v = 0.5$
 corresponds to Keplerian rotation. Furthermore, the disk is oriented
 so that the $V_0$ is always positive. Accordingly, PA varies between
 0 and $360^\circ$, while $i$ is constrained between $-90^\circ$ and
 $90^\circ$.
 \item $T_m$ and $q_m$, the temperature value at a reference
 radius $R_T$ and its exponent.
 \item $dV$, the local line width, and its exponent $e_{v}$. 
 \item $\Sigma_m$, the molecular surface density at a radius $R_\Sigma$
 and its exponent $p_m$
 \item $R_\mathrm{out}$, the outer radius of the emission, and
 $R_\mathrm{in}$, the inner radius.
 \item $h_m$, the scale height of the molecular distribution at a
 radius $R_h$, and its exponent $e_h$: it is assumed that the density
 distribution is Gaussian, with
 \begin{equation}\label{eq:scale} n(r,z) = \frac{\Sigma(r)}{h(r) \sqrt{\pi}}
 \exp\left[-\left(z/h(r)\right)^2\right] \end{equation}
 (note that with this definition, $e_h < 0$ in realistic disks),
 \end{itemize}
which gives a grand total of 17 parameters to describe a pure
spectral line emission.

All these parameters can be constrained for each observed 
line under the above assumption of power laws. This comes from two 
specific properties of protoplanetary disks: i) the rapid decrease in 
surface density with radius, and ii) the known kinematic pattern. 
In particular, we can derive both the temperature law ($T_m,q_m$) and
the surface density law ($\Sigma_m,p_m$) when there is a region
of optically thick emission (in the inner parts) while the outer
parts is optically thin.

\vspace{0.5cm}
When dust emission is not negligible, it can also be accounted for in
the emission process. Again assuming simple power laws, this adds
up six new parameters: two for the dust temperature, two for
the dust surface density, and the inner and outer radii 
of the dust distribution. The absolute value of the dust
surface density is degenerate with that of the dust absorption coefficient.
Surface density and temperature may also be degenerate if the dust
emission is optically thin and in the Rayleigh Jeans regime. Because dust emission
is in general weak compared to the observed spectral lines, an
inaccurate model of the dust will have limited effects.

Power laws are a good approximation for the velocity, temperature
\citep[see, e.g.,][]{Chiang+Goldreich_1997}, and thus to the scale height 
prescription. For molecular surface density, the approximation
may be less good because of chemical effects.

We refer to \citet{Pietu+Dutrey+Guilloteau+etal_2007} for a more 
thorough discussion of the interpretation of the model parameters. 
We recall, however, that the temperature derived in this way for a 
molecule is the excitation temperature of the transition, and that the 
surface density is derived assuming that this temperature also represents 
the rotation temperature of the rotational level population. 

From the disk model, an output data cube representing the spatial
distribution of the emitted radiation as a function of velocity is
generated by ray-tracing. From this model data cube,  DiskFit
computes the model visibilities on the same $(u,v)$ sampling
as the observed data, and derives the corresponding $\chi^2$:
\begin{equation}
\chi^2 = \Sigma_i \left(M(u_i,v_i) - O(u_i,v_i)\right)^2/W_i 
\end{equation}
where $M$ is the model visibility at the ($u_i,v_i$) Fourier plane coordinate,
$O$ the observed visibility, and $W_i$ is the visibility weight, computed
from the observed system temperature, antenna efficiency, integration
time, and correlation losses.

A Levenberg-Marquardt method (with adaptive steps adjusted according
to the estimated parameter error bars) was then used to minimize the
$\chi^2$ function upon the variable parameters. 

Error bars were computed from the covariance matrix. As described by 
\citet{Pietu+Dutrey+Guilloteau+etal_2007}, although there are many 
parameters in the model, they are in general well decoupled if
the angular resolution is sufficient. Thus the covariance matrix is 
well behaved, but asymmetric error bars are not handled (asymmetric 
error bars often occur for the outer radius, and even lead to lower 
limits only in case of insufficient sensitivity).

\paragraph{Broken power laws}

The basic power-law model above is insufficient to represent the 
emission from the GG Tau A disk because of strong and nonmonotonic 
radial variations of the line brightness in CO and $^{13}$CO.

Instead of representing the whole emission by unique temperature and 
surface density power laws over the whole extent of the disk, we therefore 
broke them into multiple power laws, each applying to different annuli. 
Such a broken power law is fully characterized by the values of the 
temperature and surface density at the knot radii, that is, the radii that 
separate consecutive annuli. Based on the knot position, the power law 
exponent can be derived from the ratio of values at consecutive knots. 
For the innermost annulus (between the inner radius and the first knot) 
and outermost annulus (between the last knot and the outer radius), we 
simply assumed the same exponent as in their respective neighbors.

This representation gives us more flexibility in the shape of the 
distribution. However, the finite spatial resolution (even accounting 
for the super-resolution provided by the Keplerian nature of the 
rotation), and sensitivity problems limit the possible number of 
knots.  In practice, we were able to use four or five knots to represent the narrow 
dense ring and the shallower outer disk in CO and other molecules.

This finer radial profile representation also prevents us 
from determining both the temperature and the surface density in each annulus, 
because these two quantities are degenerate if the line is optically thin 
(unless the annulus is very wide)\footnote{In the optically thick case, the temperature is well 
constrained, but the surface density can only be constrained from the optically thin line wings.}.

We therefore used the CO(3--2) line to derive the temperature, and
used this temperature law as fixed input parameters to derive
molecular surface densities of $\tco$(3--2) and $\cdo$(3--2).

\subsection{Best-fit model}

A global best-fit model of all three data sets cannot be performed.
We used a specific method to derive our best-fit parameters 
as described below.

\paragraph{Verification process}
In our final fit, we used fixed values for a number of parameters
in the disk model, such as the geometric parameters or parameters
related to dust emission. 
However, in the course of our study, we actually fitted many of these 
parameters, and obtained independent best fit values and errorbars from 
each data set (CO, $^{13}$CO and C$^{18}$O). The fixed value
adopted in the last step is within the error bars of all these
determinations. These parameters are marked ``verified'' in Table
\ref{tab:diskfit}

\paragraph{Geometric parameters}
All data sets were recentered on the dust ring center.

We verified by fitting that the geometric parameters were
consistent with values derived from previous studies. In particular, 
we verified that the ring center position $(X_0=0,Y_0=0)$ was also consistent
with the kinematic center of the Keplerian rotating disk.

The typical errors on these parameters ($\pm 0.01''$  for the position, 
$\pm(1^\circ-2^\circ)$ for PA and $i$, $\pm 0.03$ km\,s$^{-1}$ for 
$V_\mathrm{sys}$ and $V_0$) are far too small to affect the derived 
temperatures and surface densities in any substantial way.
Similarly, the small difference between the rotation velocity 
derived in Sec.\ref{sub:disk:kine} and the adopted value
has no significant effect.

\paragraph{Dust model}
The dust properties and dust temperature law were adopted
from \citet{Dutrey+DiFolco+Guilloteau_2014}. Only the dust surface density
was adjusted to compensate to first order flux calibration
errors. Although the model is not perfect (in particular,
it does not represent the $\sim 15 \%$ azimuthal brightness
variations), the residuals are small enough to have negligible
influence on the results derived for the observed molecules.

\paragraph{Temperature law}
The temperature law was derived from the fit to the CO data, and
used for other molecules as fixed input parameters. To better
model the ensemble, we assumed the CO column density (which is
not well constrained by the CO data because of the high
optical depth) is equal to 70 times the $^{13}$CO column density.

\paragraph{Scale height}
We assumed the scale height exponent was $e_h=-1$, that is,
$h(r) = H_m (r/r_h)$. The scale height was fit independently
for CO and $^{13}$CO data, leading to a consistent value
of $23$\,au at $r_h = 200$\,au, which was used as a fixed parameter
in the final fit for all spectral lines.

\paragraph{Nominal fit}
Table \ref{tab:diskfit} summarizes the adopted fixed parameters for our 
final best fit. Because the coupling between these parameters and
the fitted ones (temperatures and surface densities) is weak,
fixing these parameters does not affect the derived values
and errorbars of the fitted parameters.

\begin{figure*}
    \centering
    \includegraphics[width=17.0cm]{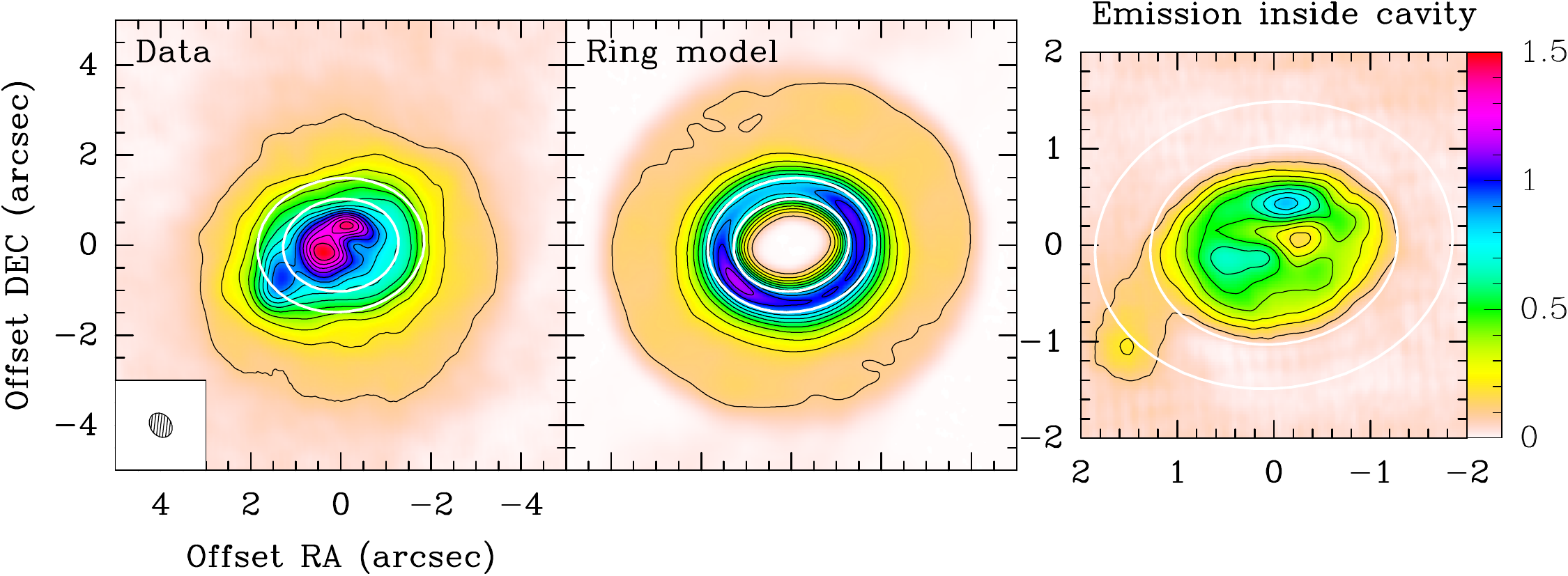}
     \hspace{0.07cm}
  \large{\rotatebox{90}{\hspace{2.5cm}Jy\,beam$^{-1}$\,km\,s$^{-1}$}}
    \includegraphics[width=17.0cm]{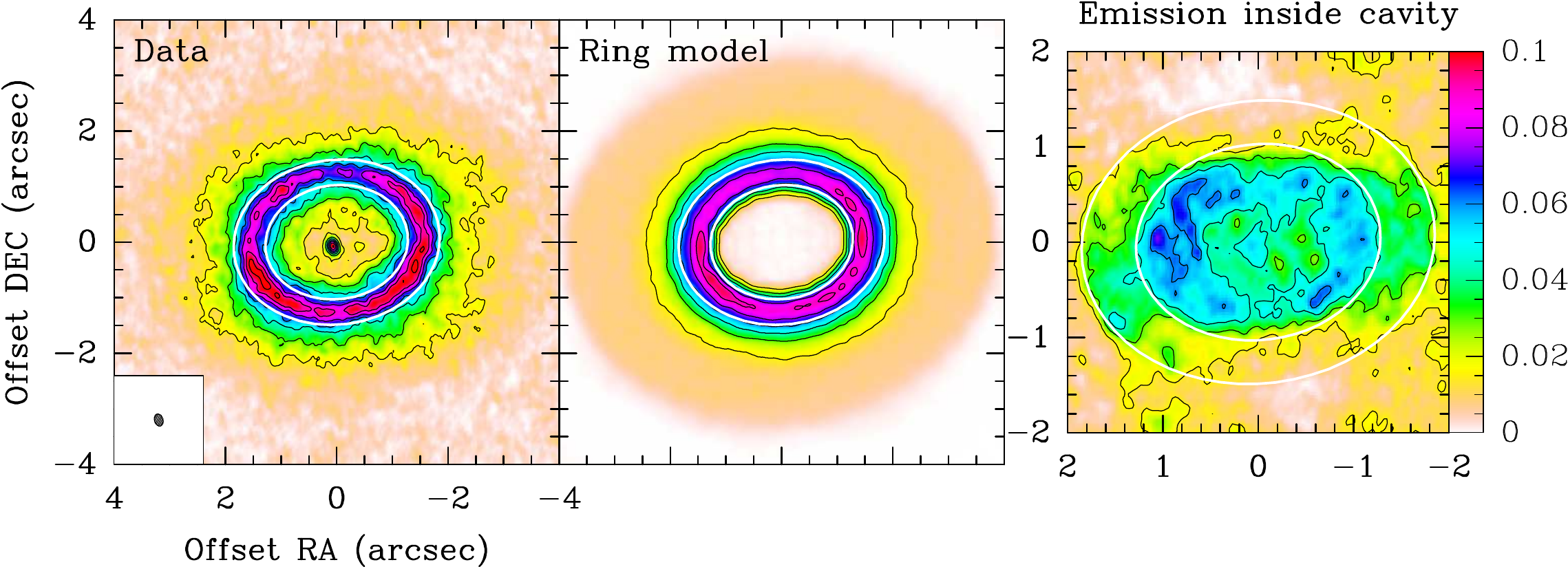}
     \hspace{0.07cm}
  \large{\rotatebox{90}{\hspace{2.5cm}Jy\,beam$^{-1}$\,km\,s$^{-1}$}}
    \includegraphics[width=17.0cm]{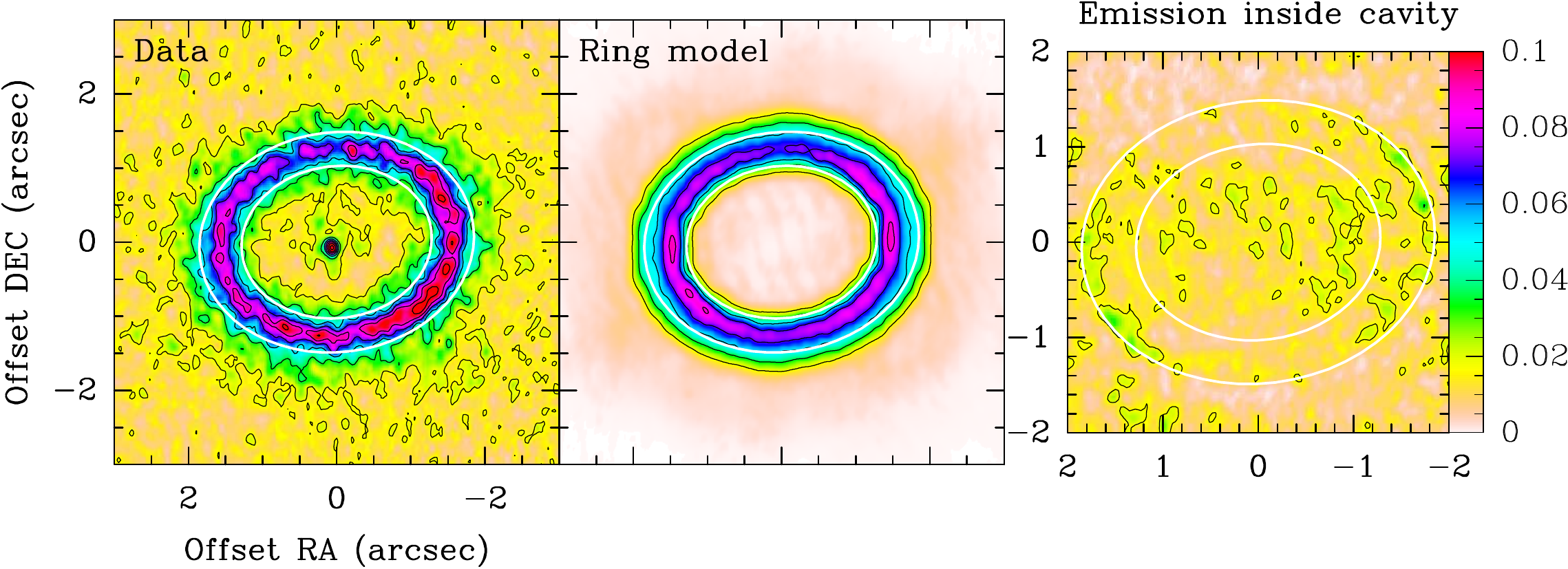}
         \hspace{0.07cm}
  \large{\rotatebox{90}{\hspace{2.5cm}Jy\,beam$^{-1}$\,km\,s$^{-1}$}
}
\caption{From top to bottom, $\dco$(3--2), $\tco$(3--2) and C$^{18}$O(3--2).  
{\it{Left panel:}} 
Integrated intensity map. {\it{Middle panel:}} Intensity map of ring and outer disk best-fit model.
{\it{Right panel:}} Residual emission inside the cavity (original data minus best-fit model).}
   \label{fig:modelco}
\end{figure*}

\paragraph{Model and residual}
Figure \ref{fig:modelco} displays the integrated intensity maps
derived from Figs.\ref{fig:13co32new}-\ref{fig:c18o32new} and from the best-fit
model, as well as that of the residuals, which are dominated by emission
in the cavity. Note that the continuum emission from Aa was
removed from these residuals.

\begin{table}[h!]   
\label{tab:diskfit} 
\caption{Fitting parameters}
\begin{center}
 \begin{tabular}{lcll}
\hline
\multicolumn{4}{c}{Geometric parameters}\\
\hline
Parameter &  Value &   & Status  \\
\hline
$(x_0,y_0)$          &  (0,0)  & Center of the dust ring & Verified\\
$PA (^{\circ})$      &   $7$   & PA of the disk rotation axis & Verified\\
$i (^{\circ}) $      &  $-35$  & Inclination & Verified \\
$V_\mathrm{LSR}$ (km\,s$^{-1})$    &  6.40   & Systemic velocity & Verified \\
$V_0$ (km\,s$^{-1})$ & $3.37$  & Keplerian rotation velocity at 100 au & Verified \\
$dV$ (km\,s$^{-1})$  & $0.3$   & Local line width & Fixed \\
$H_0$                & $24$    & Scale height at 200 au & Verified \\
\hline
\multicolumn{4}{c}{Dust ring parameters}\\
\hline
$R_\mathrm{in}$ (au) & 193 & Inner radius & fixed \\
$R_\mathrm{out}$ (au) & 285 & Outer radius &  fixed \\  
K$_{\nu}$ (cm$^{2}/g$) & $ 0.02 \times (\nu/230\mathrm{GHz})^{+1}$  & Abs. coefficient &  fixed \\    
$T(r)$ (K) & $14  \times (r/200\mathrm{\,au})^{-1}$ & Temperature  & fixed  \\
$\Sigma(\mathrm{H}_2)(r)$ (cm$^{-2}$) & 5.6 $10^{24} \times (r/200\mathrm{\,au})^{-1.4}$ & Surface density & Fitted\\
\hline                           
\multicolumn{4}{c}{Line parameters}\\
\hline
$R_\mathrm{in}$ (au)  & 169& Inner radius & Verified  \\
$R_\mathrm{knots} $ (au) & see results   & Knot positions & Fixed \\
$T(r)$ (K) &  see results  & Temperature law  & fitted from CO  \\
$\Sigma(\mathrm(X))(r)$ (cm$^{-2}$) & see results & Molecule X surface density & Fitted\\
$R_\mathrm{out}(X)$ & see results   & Outer radius for molecule X  & Fitted \\
\hline
\end{tabular}
\end{center}
\vspace{-0.5cm}
\tablefoot{Fixed values are taken from \citet{Dutrey+DiFolco+Guilloteau_2014}. Verified values
were used as fixed parameters in the last fitting step, but as free parameters in 
intermediate fits to verify their impact. }
\end{table}

\end{appendix}
\end{document}